\documentclass[11pt,a4paper]{article}
\pdfoutput=1
\usepackage{jheppub}
\usepackage{mysty}
\newcommand*{\fr}{\mathfrak{r}}

\title{Conductivities and excitations of a holographic flavour brane Weyl semimetal}
\author[1]{Haruki Furukawa,}
\author[2]{Sacha Ployet,}
\author[3]{and Ronnie Rodgers}
\affiliation[1]{Department of Physics, University of Illinois, Chicago, Illinois 60607, USA}
\affiliation[2]{Universit\'{e} Paris-Saclay, 3 rue Joliot Curie, 91190 Gif-sur-Yvette, France}
\affiliation[3]{Nordita, Stockholm University and KTH Royal Institute of Technology,
Hannes Alfv\'{e}ns v\"{a}g 12, SE-106 91 Stockholm, Sweden}
\emailAdd{hfuruk2@uic.edu}
\emailAdd{sacha.ployet@universite-paris-saclay.fr}
\emailAdd{ronnie.rodgers@su.se}
\abstract{We compute the electrical conductivities at non-zero frequency in a top-down holographic model of a Weyl semimetal, consisting of $\mathcal{N}=4$ supersymmetric $\mathrm{SU}(N_c)$ Yang--Mills theory coupled to $\mathcal{N}=2$ hypermultiplets with mass $m$, subject to an applied axial vector field $b$. The model exhibits a first-order phase transition between a Weyl semimetal phase at small $m/b$ and an insulating phase at large $m/b$. The conductivities develop peaks and troughs as functions of real frequency at low temperatures and for $m/b$ close to the phase transition. We compute the poles of the conductivities as functions of complex frequency, finding poles with small imaginary part at low temperatures and close to the phase transition.}

\begin{document}

\maketitle

\section{Introduction}
\label{sec:intro}

Weyl semimetals (WSMs) are a class of materials whose electronic bands intersect at an even number of points at or near the Fermi energy. The points in momentum space where the bands meet are called \emph{Weyl nodes}. The defining property of WSMs is that the low energy excitations of electrons around the Weyl nodes are well described by an effective theory of relativistic Weyl fermions, with the Fermi velocity \(v_f\) playing the role of the speed of light. Half of the Weyl nodes are left-handed---meaning that they host left-handed Weyl fermions---while the other half are right-handed. Reviews of WSM physics may be found in refs.~\cite{PhysRevB.84.235126,annurev:/content/journals/10.1146/annurev-conmatphys-031016-025458,annurev:/content/journals/10.1146/annurev-conmatphys-033117-054129}, for example.

One of the key features of WSMs is that they are topological materials; the degeneracy of the bands at the Weyl nodes is protected against small perturbations by non-trivial topology of the Berry curvature. The only way for Weyl nodes to disappear is for a left-handed and a right-handed node to be brought together in momentum space, where they can annihilate each another. A consequence of this topological nature is that WSMs with boundaries exhibit current carrying surface states called Fermi arcs~\cite{PhysRevB.83.205101}.

A WSM must necessarily break either inversion symmetry or time-reversal invariance; if both inversions and time-reversal are preserved then Weyl fermions must appear in degenerate pairs of opposite handedness, i.e. the low energy excitations must be Dirac fermions, not Weyl fermions. Known examples of WSMs include TaAs~\cite{Xu_2015,Lu_2015,Lv_2015,Lv_2015b}, which breaks inversions, and Co\(_3\)Sn\(_2\)S\(_2\), which breaks time-reversal~\cite{Morali_2019,Liu_2019}.

The vanishing or small density of states at the Fermi energy in a WSM causes the Coulomb interactions between electrons to be weakly screened. This makes it possible, but does not guarantee, that the electrons in a given WSM may be strongly coupled, leading to a breakdown in perturbation theory. For instance, the inversion breaking WSM Ce\(_3\)Bi\(_4\)Pd\(_3\) exhibits evidence of strong coupling~\cite{PhysRevLett.118.246601,Kirschbaum_2024}.

One tool for modelling strongly coupled systems is the AdS/CFT correspondence, also known as holography~\cite{Maldacena:1997re,Gubser:1998bc,Witten:1998qj}, which relates strongly coupled quantum field theories (QFTs) to weakly coupled gravitational theories. A number of holographic models of strongly coupled WSMs have been constructed~\cite{Gursoy:2012ie,Jacobs:2015fiv,Landsteiner:2015lsa,Landsteiner:2015pdh,Copetti:2016ewq,Liu:2018spp,Juricic:2020sgg,BitaghsirFadafan:2020lkh,Hashimoto:2016ize,Kinoshita:2017uch}, see ref.~\cite{Landsteiner:2019kxb} for a review, and various properties and phenomena have been studied within these models, including the existence of edge currents arising from surface states~\cite{Ammon:2016mwa}, an odd viscosity connected to the breaking of time reversal~\cite{Landsteiner:2016stv}, and the effect of dislocations~\cite{Juricic:2024tbe}. Existing holographic WSM models are of the type that breaks time-reversal, with the construction of a holographic inversion-breaking WSM remaining an open challenge.

One of the key advantages of holography is that it is well suited to the study of transport phenomena and hydrodynamics. The AC conductivities of the model of refs.~\cite{Landsteiner:2015lsa,Landsteiner:2015pdh} have been computed~\cite{Grignani:2016wyz}, and its hydrodynamic modes studied in ref.~\cite{Rai:2024bnr}.

Many holographic WSM models are inspired by a simple free quantum field theory (QFT) model, consisting of a Dirac fermion \(\y\) with mass \(m\) coupled to an external axial vector field \(\vec{A}_5 = (0,0,b/2)\) for some constant \(b\), where without loss of generality we have chosen to orient \(\vec{A}_5\) in the \(z \equiv x^3\) direction. The Lagrangian density of this model is~\cite{Colladay:1998fq,Grushin:2012mt}
\begin{equation} \label{eq:free_wsm_lagrangian}
    \mathcal{L} = i \bar{\y}\le(i \g^\m \p_\m - m + \frac{b}{2}\g^3 \g^5 \ri) \y.
\end{equation}
The single particle Hamiltonian that follows from equation~\eqref{eq:free_wsm_lagrangian} has four eigenvalues which are functions of momentum \(\vec{k}\). When the axial vector field is sufficiently strong, \(b^2 > 4 m^2\), two of these eigenvalues meet at \(\vec{k} = (0,0,\pm\sqrt{(b/2)^2 - m^2})\). These are the two Weyl nodes. When \(b^2 < 4 m^2\) the Hamiltonian is gapped and the system is insulating. The transition at \(b^2 = 4m^2\) between the WSM and insulating phases occurs via the mechanism described above, in which two Weyl nodes are brought together in momentum space and annihilate each other.

One of the striking features of the model in equation~\eqref{eq:free_wsm_lagrangian} is that the separation of the Weyl nodes in momentum space induces an axial anomaly, with observable consequences for transport. If the model is in the WSM phase, \(b^2 < 4 m^2\), then thanks to the axial anomaly an electric field \(\vec{E}\) applied in a direction orthogonal to the \(\vec{A}_5\) will produce a vector current \(\vev{\vec{J}}\) in the direction orthogonal to both \(\vec{E}\) and \(\vec{A}_5\), in addition to the usual current parallel to \(\vec{E}\). Concretely, a uniform applied electric field produces a current
\begin{equation} \label{eq:current_orthogonal_to_b}
    \vev{\vec{J}} =\begin{pmatrix}
        \s_{xx} & \s_{xy}& 0\\
        - \s_{xy} & \s_{yy} & 0\\
        0 & 0 & \s_{zz}
    \end{pmatrix} \vec{E},
\end{equation}
where rotational symmetry implies that \(\s_{xx} = \s_{yy}\), and the Hall conductivity, which is non-zero in the WSM phase due to the axial anomaly, is~\cite{Landsteiner:2016led}
\begin{equation} \label{eq:free_hall_conductivity}
    \s_{xy} = \frac{\mathrm{sign}(b)}{4\pi^2} \sqrt{b^2 - 4m^2} \, \q(|b| - 2m),
\end{equation}
where \(\q\) is the Heaviside step function.

Roughly speaking, most existing holographic WSM models use gravity to describe a QFT with similar features to the free model in equation~\eqref{eq:free_wsm_lagrangian} but with the addition of strong interactions. Many such models are \emph{bottom-up}, meaning that they are based on a gravitational theory for which the dual QFT (if it exists) is not known. Other holographic WSM models are \emph{top-down}~\cite{Copetti:2016ewq,BitaghsirFadafan:2020lkh,Hashimoto:2016ize,Kinoshita:2017uch}, meaning that they originate from a supergravity (SUGRA) or string theory construction, allowing the dual QFT to be determined.

One top-down holographic WSM model is that of ref.~\cite{BitaghsirFadafan:2020lkh}, reviewed in section~\ref{sec:review}. This model originates from an intersection of \(N_c\) D3-branes and \(N_f\) D7-branes in type IIB SUGRA. The holographic dual QFT is a strongly coupled \(\cN=4\) supersymmetric \(\SU(N_c)\) Yang--Mills theory with 't Hooft coupling \(\l\), coupled to \(N_f\) \(\cN=2\) hypermultiplets~\cite{Karch:2002sh}. The Lagrangian density for this QFT includes terms precisely of the form in equation~\eqref{eq:free_wsm_lagrangian}, describing the coupling of a Dirac fermion \(\y\) with mass \(m\) (belonging to the hypermultiplets) to a background axial field proportional to a parameter \(b\). There are also many other terms in the Lagrangian, describing the other matter fields and their interactions. We think of \(\y\) as describing the electrons in the WSM, with the other fields mediating the strong inter-electron interactions.  

We will refer to the model of ref.~\cite{BitaghsirFadafan:2020lkh} as the (holographic) flavour brane WSM model. We note that the flavour brane model is distinct from the model of refs.~\cite{Hashimoto:2016ize,Kinoshita:2017uch}, which is also based on a D3/D7 intersection but realises WSM physics in a different way, through Floquet states arising from a rotating applied electric field.\footnote{See refs.~\cite{Garbayo:2020dmh,Berenguer:2022act} for further discussion of Floquet states dual to D-brane intersections.} This is a holographic realisation of a proposal from ref.~\cite{wang2013floquetweylsemimetalinduced}.

\begin{figure}[!htbp]
    \begin{center}
        \includegraphics{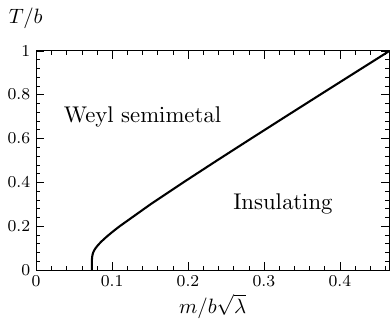}
    \end{center}
    \caption{Phase diagram of the flavour brane WSM model, in the plane of temperature \(T\) versus hypermultiplet mass \(m\) in units of the applied axial vector field \(b\). The black curve shows the location of a first-order phase transition. To the left of the phase transition, at small mass, there is a Weyl semimetal phase. To the right of the phase transition, at large mass, there is an insulating phase. At zero temperature the phase transition occurs at \(m \approx 0.0733 \, b \sqrt{\l}\). Figure adapted from ref.~\cite{BitaghsirFadafan:2020lkh}.}
    \label{fig:phase_diagram}
\end{figure}

Ref.~\cite{BitaghsirFadafan:2020lkh} computed the phase diagram of the flavour brane WSM model, finding the result plotted in figure~\ref{fig:phase_diagram}. As for the free QFT model in equation~\eqref{eq:free_wsm_lagrangian}, the system undergoes a phase transition. For sufficiently small \(m/b\) the model exhibits a WSM phase, while for large \(m/b\) one finds an insulating phase. There is a first-order phase transition at a critical value of \(m/b\) that depends depends on the gauge theory's 't Hooft coupling as \(\l^{-1/2}\), and also depends on temperature. At zero temperature the phase transition occurs at \(m \approx 0.0733 \, b \sqrt{\l}\).\footnote{Ref.~\cite{Evans:2024ilx} showed that this first-order phase transition persists at non-zero magnetic field \(\vec{B}\) and also studied the dynamics of bubble walls at non-zero \(\vec{B}\).}

\begin{figure}[!htbp]
    \begin{subfigure}{0.5\textwidth}
        \includegraphics[width=\textwidth]{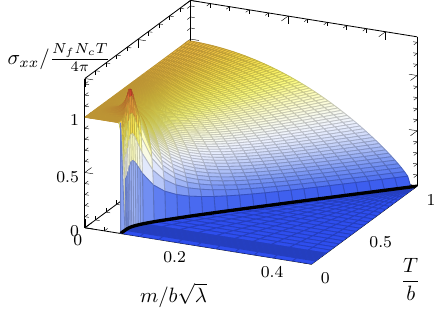}
        \caption{Longitudinal conductivity}
        \label{fig:old_dc_conductivities_xx}
    \end{subfigure}\begin{subfigure}{0.5\textwidth}
        \includegraphics[width=\textwidth]{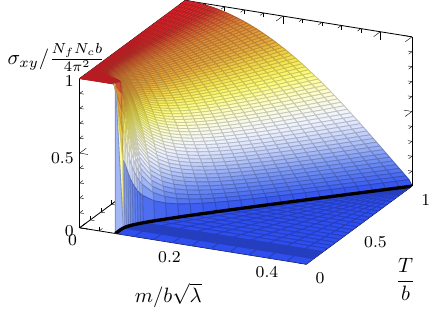}
        \caption{Hall conductivity}
    \end{subfigure}
    \caption{
        Longitudinal and Hall conductivities in the holographic D7-brane Weyl semimetal model. The solid black curve shows the location of the phase transition plotted in figure~\ref{fig:phase_diagram}. The conductivities vanish in the insulating phase and are non-zero in the Weyl semimetal phase. Figure adapted from ref.~\cite{BitaghsirFadafan:2020lkh}.
    }
    \label{fig:old_dc_conductivities}
\end{figure}

The longitudinal and Hall conductivities \(\s_{xx}\) and \(\s_{xy}\) appearing in equation~\eqref{eq:current_orthogonal_to_b} were also computed in ref.~\cite{BitaghsirFadafan:2020lkh}. The results are plotted in figure~\ref{fig:old_dc_conductivities}. They are functions of the dimensionless ratios \(m/b\sqrt{\l}\) and \(T/b\). They vanish in the insulating phase, of course, and are non-zero in the WSM phase. For fixed \(T/b\) the conductivities are typically largest at small \(m/b\sqrt{\l}\), and decrease as \(m/b\sqrt{\l}\) is increased towards the phase transition. The exception is \(\s_{xx}\) at small \(T/b\), which exhibits a peak close to the phase transition.

In this work, we extend the results of ref.~\cite{BitaghsirFadafan:2020lkh} by computing the AC conductivities, which determine the current produced by a time-dependent electric field. We will also complete the results plotted in figure~\ref{fig:old_dc_conductivities} by computing the DC value of \(\s_{zz}\). We will find that the AC conductivities exhibit features (peaks and troughs) as functions of frequency for zero or small \(T/b\) and values of \(m/b\sqrt{\l}\) close to the phase transition, in a manner somewhat reminiscent of the DC \(\s_{xx}\) as a function of temperature plotted in figure~\ref{fig:old_dc_conductivities_xx}. These features are suggestive of poles in the conductivities close to the real axis in the complex frequency plane. This motivates us to compute these poles at zero temperature, which are holographically dual to quasinormal modes of the gravitational system~\cite{Birmingham:2001pj,Son:2002sd,Kovtun:2005ev}.

This paper is organised as follows. In section~\ref{sec:review} we review the holographic flavour brane WSM model, focusing on the features needed to understand our calculations. In section~\ref{sec:conductivity} we present our results for the conductivities. We then present the poles in the conductivities in section~\ref{sec:qnm}. We close with our conclusions and outlook for the future in section~\ref{sec:discussion}. Some additional calculational details and plots are given in the appendices.

The numerical results presented in this article, along with a LaTeX file containing PGFPlots code to reproduce the plots, may be downloaded from the accompanying data release~\cite{data_realease}.

\section{Review: The holographic flavour brane Weyl semimetal}
\label{sec:review}

In this section we review the holographic flavour brane WSM model of ref.~\cite{BitaghsirFadafan:2020lkh}. Readers familiar with the model may safely skip this section, referring back as needed for definitions and notation.

Four-dimensional \(\cN=4\) supersymmetric Yang--Mills (SYM) theory with gauge group \(\SU(N_c)\) is holographically dual to type IIB supergravity (SUGRA) on \(\ads[5] \times \sph[5]\), which arises as the near-horizon limit of the solution describing a flat stack of \(N_c\) D3-branes~\cite{Maldacena:1997re,Gubser:1998bc,Witten:1998qj}. The \(\ads[5] \times \sph[5]\) solution has a constant dilaton and a non-trivial metric and Ramond--Ramond four-form \(C_4\), with all other fields of type IIB SUGRA vanishing. We will write the metric and four-form \(C_4\) of this solution in the gauge\footnote{Compared to ref.~\cite{BitaghsirFadafan:2020lkh} we have rescaled the \(r\) and \(R\) coordinates by \(L^2\), which will eliminate factors of \(L\) from many of our subsequent expressions. Our \(r\) and \(R\) have units of energy.}
\begin{equation}\begin{aligned} \label{eq:ads5_cross_s5_zero_T}
    \diff s^2 &= L^2 \r^2 \le( - \diff t^2 + \diff \vec{x}^2 \ri) + \frac{L^2}{\r^2} \le( \diff r^2 + r^2 \diff s_{\sph[3]}^2 + \diff R^2 + R^2 \diff \f^2 \ri), \qquad 
    \\
    C_4 &= L^4 \r^4 \, \diff t \wedge \diff x \wedge \diff y \wedge \diff z - \frac{L^4}{\r^4} r^4 \, \diff \f \wedge \w(\sph[3]),
\end{aligned}\end{equation}
where we have defined~\(\r^2 = r^2 + R^2\), while \(L\) is the curvature radius of both the \ads[5] and \sph[5] factors, related to the Regge slope \(\a'\) and the dual \(\cN=4\) SYM theory's 't Hooft coupling \(\l\) by \(L^4 = \l \a'^2\). The directions \(t\) and 
\(\vec{x} = (x,y,z)\) are the directions parallel to the stack of D3-branes, and may be thought of as the coordinates of the dual \(\cN=4\) SYM. We have defined \(\diff s_{\sph[3]}^2\) and \(\w(\sph[3])\) as the metric and volume form respectively of a unit, round three-sphere.

If we replace the \(\ads[5]\) factor in this geometry with an \ads[5]-Schwarzschild black brane, the geometry is now holographically dual to \(\cN=4\) SYM at non-zero temperature \(T\) equal to the Hawking temperature of the black brane~\cite{Gubser:1996de,Witten:1998zw}. The new geometry is
\begin{equation}\begin{aligned} \label{eq:ads5_cross_s5}
    \diff s^2 &= L^2 \r^2 \le( - \frac{f(\r)^2}{h(\r)} \diff t^2 + h(\r) \diff \vec{x}^2 \ri) + \frac{L^2}{\r^2} \le( \diff r^2 + r^2 \diff s_{\sph[3]}^2 + \diff R^2 + R^2 \diff \f^2 \ri),
    \\
    C_4 &= L^4 \r^4 h(\r)^2 \, \diff t \wedge \diff x \wedge \diff y \wedge \diff z - \frac{L^4}{\r^4} r^4 \, \diff \f \wedge \w(\sph[3]),
\end{aligned}\end{equation}
where
\begin{equation}
    \qquad
    f(\r)  = 1 - \frac{\r_h^4}{\r^4},
    \qquad
    h(\r)  = 1 + \frac{\r_h^4}{\r^4}.
\end{equation}
The horizon of the \ads[5]-Schwarzschild black brane is located at \(\r = \r_h\). Its Hawking temperature, or equivalently the temperature of the dual \(\cN=4\) SYM, is
\begin{equation}
    T = \frac{\sqrt{2}}{\pi} \r_H.
\end{equation}
The \(\ads[5]\times\sph[5]\) solution~\eqref{eq:ads5_cross_s5_zero_T} is straightforwardly recovered from equation~\eqref{eq:ads5_cross_s5} by setting \(T=0\) (equivalently, \(\r_h = 0\)), which sets \(f(\r) = h(\r)=1\).

\begin{table}
    \begin{center}\begin{tabular}{c | c c c c  c  c  c c  c c}
       & \(t\)\! & \(x\) \! & \(y\) & \(z\) & \(r\) & \multicolumn{3}{c}{\(\sph[3]\)} & \(R\) & \(\f\)
       \\
       \hline
       D3 & \(\times\) & \(\times\) & \(\times\) & \(\times\) & \(\cdot\) & \(\cdot\) & \(\cdot\) & \(\cdot\) & \(\cdot\) & \(\cdot\)
       \\
       D7 & \(\times\) & \(\times\) & \(\times\) & \(\times\) & \(\times\) & \(\times\) & \(\times\) & \(\times\) & \(\cdot\) & \(\cdot\)
   \end{tabular}\end{center}
    \caption{In the D3/D7 WSM model, there are \(N_c\) D3-branes spanning the four directions \((t,x,y,z)\), as indicated by the crosses in the first row. The D3-branes produce the geometry in equation~\eqref{eq:ads5_cross_s5}. There are also \(N_f\) probe D7-branes, spanning \((t,x,y,z,r)\) and an \sph[3], as indicated by the crosses in the second row. The two directions orthogonal to both the D3- and D7-branes form a plane, on which we use polar coordinates \((R,\f)\). WSM physics is realised when \(\f\) is linear in one of the spatial directions parallel to the D3-branes, \(\f \propto z\).}
    \label{tab:intersection}
\end{table}

The next step in constructing the D3/D7 WSM model is to introduce \(N_f\) coincident D7-branes into this geometry, spanning \(\xi = (t,\vec{x},r,\sph[3])\). The intersection is summarised in table~\ref{tab:intersection}. These D7-branes are dual to \(N_f\) \(\cN=2\) hypermultiplets coupled to \(\cN=4\) SYM~\cite{Karch:2002sh}. We will always work in the probe limit \(N_f \ll N_c\), in which we can neglect the back-reaction of the D7-branes on the metric and \(C_4\).

The terms of the D7-brane action that are relevant for what follows are
\begin{equation} \label{eq:D7_action}
    S_\mathrm{D7} = - N_f T_\mathrm{D7} \int d^8 \xi \sqrt{-\det (P[g] + 2 \pi \a' F)}
    + 2 \pi^2 \a'^2 N_f T_\mathrm{D7} \int P[C_4] \wedge F \wedge F,
\end{equation}
where \(T_\mathrm{D7}\) is the tension of a single D7-brane, \(F = \diff A\) is the field strength for a \(\mathrm{U}(1)\) world-volume gauge field \(A\), and \(P[g]\) and \(P[C_4]\) are the pullback of the metric and four-form in equation~\eqref{eq:ads5_cross_s5}. When measured in units of the \ads\ radius \(L\), the D7-brane tension is related to the 't Hooft coupling \(\l\) and number of colours \(N_c\) in the dual QFT by \(T_\mathrm{D7} L^8 = \l N_c / (32 \pi^6)\).

In addition to the gauge field \(A\), the D7-branes have two world-volume scalars \(R\) and \(\f\), which parameterise the embedding of the D7-branes in the bulk spacetime. To construct the D7-brane embeddings dual to a WSM, we make the ansatz that \(A=0\) and
\begin{equation} \label{eq:D7_ansatz}
    R = R(r), \qquad \f = b z.
\end{equation}
The ansatz for \(\f\) corresponds to the coupling of the fermions in the hypermultiplet to a constant axial vector field of the form in equation~\eqref{eq:free_wsm_lagrangian}, thus generating an axial anomaly and creating the possibility of realising WSM physics~\cite{Hoyos:2011us,Kharzeev:2011rw,Bu:2018trt,BitaghsirFadafan:2020lkh}. With this ansatz, the induced metric on the D7-branes, \(\diff s_{\mathrm{D7}}^2 = P[g]_{ab} \diff \xi^a \diff \xi^b\), is
\begin{equation}\begin{aligned} \label{eq:induced_metric}
    \diff s_\mathrm{D7}^2 
    &= L^2 \r^2 \le[ - \frac{f(\r)^2}{h(\r)} \diff t^2 + h(\r) \le( \diff x^2 + \diff y^2 \ri) + \le(h(\r) + \frac{b^2 R^2}{\r^4} \ri) \diff z^2\ri]
    \\
    &\hspace{7cm} + \frac{L^2}{\r^2} \le[(1+R'^2) \diff r^2 + r^2 \diff s_{\sph[3]}^2 \ri],
\end{aligned}\end{equation}

After substituting the determinant of the induced metric~\eqref{eq:induced_metric} into the D7-brane action~\eqref{eq:D7_action}, one can derive the Euler--Lagrange equation for \(R(r)\),\footnote{The ansatz also solves the Euler--Lagrange equation for \(\f\).}
\begin{equation}\begin{aligned} \label{eq:embedding_equation}
    \frac{R''}{1 + R'^2} +& \frac{8 r R + (3 R^2 - 5 r^2) R'}{r \r^2}
    - \frac{b^2 R}{\r^4 + \r_h^4 + b^2 R^2}
    \\
    &\hspace{2cm}+ 2 \r^2 r^2 \le(\frac{R}{r} \ri)' \frac{4 \r^4 (\r^4 + \r_h^4)  + b^2 (3 \r^4 + \r_h^4) R^2}{(\r^8 - \r_h^8)(\r^4 + \r_h^4 + b^2 R^2)} = 0,
\end{aligned}\end{equation}
where primes denote derivatives with respect to \(r\), and we recall the definition \(\r^2 = r^2 +R^2\). Near the asymptotically-AdS boundary at \(r \to \infty\), solutions to this equation behave as
\begin{equation} \label{eq:R_UV}
    R(r) = M  \le(1 - \frac{b^2}{2 r^2} \log(r/M) \ri) +  \frac{C}{r^2} + O\le(\frac{\log r}{r^4} \ri).
\end{equation}
where \(M\) and \(C\) are integration constants. The coefficient \(M\) is proportional to the asymptotic separation between the D3- and D7-branes and therefore determines the mass \(m\) of the \(\cN=2\) hypermultiplet,
\begin{equation}
    m = \frac{M \sqrt{\l}}{2 \pi}.
\end{equation}
For a given \(M\), we fix \(C\) by demanding that the full solution for \(R(r)\) is regular throughout the bulk of the spacetime. The vacuum expectation value of the scalar operator \(\cO_m\) sourced by \(m\) is determined by \(C\)~\cite{BitaghsirFadafan:2020lkh},\footnote{Compared to the analogous equation in ref.~\cite{BitaghsirFadafan:2020lkh}, our equation~\eqref{eq:scalar_vev} has no term proportional to \(\log (ML)\). This is because we have chosen to make the argument of the logarithm in the UV expansion~\eqref{eq:scalar_vev} dimensionless using \(M\) rather than \(L\), causing our \(C\) to differ from that of ref.~\cite{BitaghsirFadafan:2020lkh} by an additive factor proportional to \(\log(ML)\).}
\begin{equation} \label{eq:scalar_vev}
    \vev{\cO_m} = \frac{\sqrt{\l} N_f N_c}{8 \pi^3} \le(-2 C + \frac{b^2 M}{2} \ri).
\end{equation}

There are several classes of solution to the equation of motion for \(R(r)\), distinguished by their behaviour in the infrared, in other words by their behaviour near the black brane horizon (or at near \(r=0\) at \(T=0\)). The different classes of solution correspond to different phases in the dual field theory:

\paragraph{Weyl semimetal phase: black hole embeddings.}

The first class of solutions are \emph{black hole embeddings}~\cite{Mateos:2006nu}, in which the D7-branes meet the horizon at some \(r = r_h\), which must obey \(r_h \leq \r_h\). Near the horizon, solutions to the embedding equation take the form
\begin{equation} \label{eq:bh_embedding_ir}
    R(r) = r \sqrt{\frac{\r_h^2}{r_h^2} - 1}  + O\le((r - r_h)^2\ri)
\end{equation}
For fixed \(T\), different values of \(r_h\) correspond to different values of \(M\), and therefore different values of the dimensionless parameter \(m/b \sqrt{\l}\) in the dual QFT. Black hole embeddings describe phases with a continuous spectrum of excitations and non-zero conductivity~\cite{Hoyos-Badajoz:2006dzi,Karch:2007pd}. Ref.~\cite{BitaghsirFadafan:2020lkh} showed that non-zero \(b\) also leads to an anomalous Hall conductivity, prompting the identification of black hole embeddings with a WSM phase.

In the limit of zero temperature the horizon disappears and the black hole embeddings tend to a new class of solution in which \(R\) vanishes as \(r \to 0\). In a slight abuse of terminology, we will also refer to these embeddings as ``black hole embeddings'', even though there is no black hole present at \(T=0\). The small-\(r\) asymptotics of the \(T=0\) black hole embeddings are
\begin{equation} \label{eq:bh_embedding_ir_T0}
    R(r) = \eta \frac{e^{-b/r}}{\sqrt{r}} \le[1 + O(r^2) \ri],
\end{equation}
where \(\h\) is an integration constant. Different choices of \(\h\) correspond to different values of \(m/b \sqrt{\l}\). Embeddings with these asymptotics describe the WSM phase at zero temperature.

\paragraph{Insulating phase: Minkowski embeddings.}

The second class of solutions never touch the horizon. Instead, they reach all the way \(r=0\), with small-\(r\) asymptotics
\begin{equation} \label{eq:minkowski_ir}
    R(r) = R_0 + O(r^2),
\end{equation}
for some \(R_0 > \r_h\). For a given temperature, embeddings with different values of \(R_0\) correspond to different values of \(m/b \sqrt{\l}\). Embeddings of this type are usually called \emph{Minkowski embeddings}~\cite{Mateos:2006nu}, and describe insulating phases with a gapped, discrete spectrum~\cite{Hoyos-Badajoz:2006dzi,Karch:2007pd}. These Minkowski embeddings also exist at zero temperature, for any \(R_0 > 0\).

\paragraph{Critical embedding.} The third and final class of solution touches the horizon exactly at \(r = 0\). It has small-\(r\) asymptotics
\begin{equation} \label{eq:critical_embedding_ir}
    R(r) = \r_h + \frac{r}{\sqrt{3}} + O(r^3).
\end{equation}
There are no integration constants in this expansion, so for a given temperature such an embedding only exists for a single value of \(m/b \sqrt{\l}\). Such an embedding exists at the boundary between black hole and Minkowski embeddings, and is often called a \emph{critical embedding}~\cite{Mateos:2006nu}. The asymptotics in equation~\eqref{eq:critical_embedding_ir} remain valid at \(T \propto \r_h =0\).

\begin{figure}
        \includegraphics{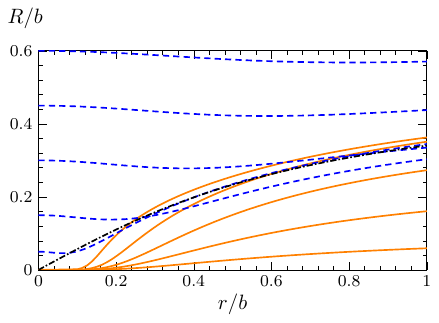}
        \includegraphics{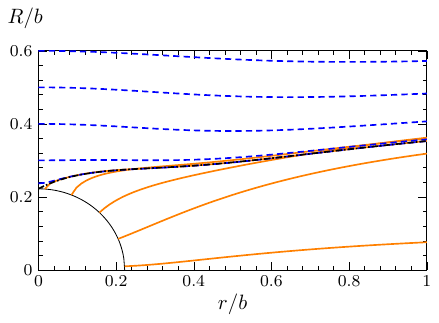}
    \caption{Example embeddings, obtained by solving equation~\eqref{eq:embedding_equation} numerically subject to the different classes of IR boundary conditions discussed in the text. The left panel shows embeddings at zero temperature \(T = 0.1\,b\), while the right panel shows embeddings at a sample non-zero temperature. The solid orange curves show black hole embeddings, the dashed blue curves show Minkowski embeddings, and the dot-dashed black curves show critical embeddings. The solid black quarter-circle in the right panel shows the location of the black brane horizon. Figure adapted from ref.~\cite{BitaghsirFadafan:2020lkh}.}
    \label{fig:embeddings}
\end{figure}

Examples of the different classes of embeddings are plotted in figure~\ref{fig:embeddings}. The solid orange curves show black hole embeddings, the dashed blue curves show Minkowski embeddings, and the dot-dashed black curves show critical embeddings. At large \(r\), the embeddings asymptote to \(R(r\to \infty) = M\). As suggested by figure~\ref{fig:embeddings}, there are values of \(M\) for which one finds both a black hole and a Minkowski embedding. Ref.~\cite{BitaghsirFadafan:2020lkh} computed the free energy of these embeddings, holographically dual to the D7-branes' on-shell action, to determine whether black hole or Minkowski embeddings are thermodynamically favoured. The result is that there is a first-order phase transition between the WSM phase and the insulating phase, leading to the phase diagram drawn in figure~\ref{fig:phase_diagram}. The WSM phase is thermodynamically favoured at small masses while the insulating phase is favoured at large masses. At zero temperature this phase transition occurs at \(m \approx 0.0733 \, b \sqrt{\l}\). The critical embedding is never thermodynamically favoured.

At zero temperature, black hole and Minkowski embeddings approach the critical embedding in the limits \(\h \to \infty\) and \(R_0 \to 0\), respectively. Similarly, at non-zero temperature, black hole and Minkowski embeddings approach the critical embedding in the limits \(r_h \to 0\) and \(R_0 \to \r_h\). Such near-critical solutions exhibit a discrete scale invariance with complex critical exponents, leading to a self-similarity of the solutions~\cite{Mateos:2006nu,Mateos:2007vn,Frolov:1998td,Frolov:2006tc,Karch:2009ph,BitaghsirFadafan:2020lkh}. This self-similarity is connected to the first-order nature of the phase transition.

\section{AC conductivities}
\label{sec:conductivity}

\subsection{On the holographic determination of the AC conductivities}

In this section we compute the AC conductivities of the Weyl semimetal phase of our holographic model. A small time-dependent applied electric field \(\vec{E}(t)\) will induce a non-zero one-point function of the vector current \(\vev{\vec{J}(t)}\). Ohm's law relates the Fourier components \(\vev{\vec{J}(\w)}\) and \(\vec{E}(\w)\) of the current and electric field as
\begin{equation}
    \vev{J_i(\w)} = \s_{ij}(\w) E_j(\w),
\end{equation}
where \(\s_{ij}(\w)\) is the AC conductivity. We will obtain the AC conductivities from the Kubo formula (see e.g. ref.~\cite{Herzog:2009xv})
\begin{equation} \label{eq:conductivity_kubo_formula}
    \s_{ij}(\w) = \frac{G_{ij}(\w,0)}{i\w},
\end{equation}
where \(G_{ij}(\w,\vec{k})\) is the retarded Green's function of the vector current components \(J^i\) and \(J^j\), computed in momentum space with frequency \(\w\) and momentum \(\vec{k}\).\footnote{
Explicitly, the retarded Green's function appearing in the Kubo formula is
\[
    G_{ij}(\w,\vec{k}) = - i \int \diff t \, \diff^3 \vec{x} \, e^{i \w t - i \vec{k}\cdot \vec{x}} \, \q(t) \, \langle [J^i(t,\vec{x}), J^j(0,0)] \rangle,
\]
where \(\q(t)\) is the Heaviside step function.
}
Retarded Green's functions are obtained holographically from linearised fluctuations of bulk fields subject to ingoing boundary conditions at the black brane horizon~\cite{Son:2002sd,Herzog:2002pc,Kaminski:2009dh}, as follows.

Adopting a radial gauge \(A_r = 0\), we consider linearised fluctuations of the field \(A_i\) holographically dual to the vector current \(J^i\) about backgrounds of the form in equation~\eqref{eq:D7_ansatz}. The \(A_i\) are the components of the D7-branes' world-volume gauge field in the spatial directions \(\vec{x} = (x,y,z)\) of the dual QFT. Since the Kubo formula~\eqref{eq:conductivity_kubo_formula} contains the Green's function at non-zero frequency but vanishing momentum, we allow for the gauge field fluctuations to depend on \(t\) and \(r\). We then Fourier transform with respect to time, writing \(A_i(t,r) = \int \frac{\diff \w}{2\pi} e^{-i \w t} A_i(\w;r)\). The equations of motion for the Fourier modes \(A_i(\w;r)\), derived from the action~\eqref{eq:D7_action}, are
\begin{equation}\begin{aligned} \label{eq:fluctuation_equations}
    \le(\frac{r^3}{\r^2} \frac{f}{\sqrt{h}} \sqrt{\frac{\r^4 h + b^2 R^2}{1 + R'^2}} A_x' \ri)'
    + \w^2 \frac{r^3}{\r^6} \frac{\sqrt{h}}{f} \sqrt{\r^4 h + b^2 R^2} \sqrt{1 + R'^2} A_x &= - i \w b \le(\frac{r^4}{\r^4} \ri)' A_y,
    \\
    \le(\frac{r^3}{\r^2} \frac{f}{\sqrt{h}} \sqrt{\frac{\r^4 h + b^2 R^2}{1 + R'^2}} A_y' \ri)'
    + \w^2 \frac{r^3}{\r^6} \frac{\sqrt{h}}{f} \sqrt{\r^4 h + b^2 R^2} \sqrt{1 + R'^2} A_y &=  i \w b \le(\frac{r^4}{\r^4} \ri)' A_x,
\end{aligned}\end{equation}
and
\begin{equation} \label{eq:A_z_eom}
    \le(\frac{r^3 \r^2  f \sqrt{h} \, A_z'}{\sqrt{(\r^4 h  +b^2 R^2)(1+R'^2)}}  \ri)'
    + \w^2 \frac{r^3}{\r^2} \frac{h^{3/2}}{f} \sqrt{\frac{1 + R'^2}{\r^4 h + b^2 R}} \, A_z = 0.
\end{equation}
For non-zero \(b\) the equations of motion for \(A_x\) and \(A_y\) are coupled. However, they may be decoupled by the introduction of two new fields \(A_\pm = A_x \pm i A_y\), the equations of motion for which are
\begin{equation} \label{eq:A_plus_minus_eom}
    \le(\frac{r^3}{\r^2} f \sqrt{h} \sqrt{\frac{\r^4 h + b^2 R^2}{1 + R'^2}} A_{\pm}' \ri)'
    + \w^2 \frac{r^3}{\r^6} \frac{\sqrt{h}}{f} \sqrt{\r^4 h + b^2 R^2} \sqrt{1 + R'^2} A_{\pm}  \pm \w b \le(\frac{r^4}{\r^4} \ri)' A_{\pm} = 0.
\end{equation}
We will use an index \(n\) to label the three components of \((A_+,A_-,A_z)\).

As reviewed in section~\ref{sec:review}, the WSM phase is dual to black hole embeddings, for which the solution for \(R(r)\) has the IR behaviour in equation~\eqref{eq:bh_embedding_ir} or~\eqref{eq:bh_embedding_ir_T0} for \(T>0\) or \(T=0\), respectively. The equations of motion for \(A_n\) then admit solutions obeying ingoing boundary conditions of the form
\begin{equation} \label{eq:ingoing_boundary_conditions}
    A_n(\w;r) \approx 
    \begin{cases}
        (r - r_h)^{-i \w/(2 \pi T)},
        & \text{for }r \to r_h \text{ at } T \neq 0,
        \\
        \dfrac{e^{i \w/r}}{\sqrt{r}},
        & \text{for }r \to 0 \text{ at } T = 0.
    \end{cases}
\end{equation}
We must impose that the fluctuations have the IR behaviour of equation~\eqref{eq:ingoing_boundary_conditions} in order to obtain the retarded Green's functions~\cite{Son:2002sd}.\footnote{The equations of motion also admit outgoing boundary conditions, which would correspond to the opposite signs in the exponents in equation~\eqref{eq:ingoing_boundary_conditions} and would be appropriate for the computation of the advanced Green's functions.}

Near the boundary at \(r \to \infty\), solutions to the equations of motion for the fluctuations behave as
\begin{equation}
    A_n(\w;r) = A_n^{(0)}(\w) \le[1 + \frac{\w^2}{4 r^2} \log(r^2/\w^2) \ri] + \frac{A_n^{(2)}(\w)}{r^2} + O(r^{-4} \log r).
\end{equation}
where \(A_n^{(0)}(\w)\) and \(A_n^{(2)}(\w)\) are integration constants, fixed by the ingoing boundary conditions (up to an overall scaling, since the equations of motion for \(A_n(\w;r)\) are linear). The integration constants determine the retarded Green's functions, which in turn determine the conductivities through the Kubo formula~\eqref{eq:conductivity_kubo_formula}. The result is that the non-zero conductivities are given by
\begin{subequations}\label{eq:sigma_formula}\begin{align}
    \s_{xx}(\w) = \s_{yy}(\w) &= \frac{N_f N_c}{4 \pi^2}  \le[
        \frac{1}{i\w} \le(  \frac{A_+^{(2)}(\w)}{A_+^{(0)}(\w)} + \frac{A_-^{(2)}(\w)}{A_-^{(0)}(\w)} \ri)
        + \frac{i\w}{2} \ri],
    \label{eq:sigma_xx_formula}
    \\
    \s_{xy}(\w) &= \frac{N_f N_c}{4 \pi^2}  \le[
        \frac{1}{\w} \le(  \frac{A_-^{(2)}(\w)}{A_-^{(0)}(\w)} - \frac{A_+^{(2)}(\w)}{A_+^{(0)}(\w)} \ri)
        + b \ri],
    \label{eq:sigma_xy_formula}
    \\
    \s_{zz}(\w) &= \frac{N_f N_c}{4 \pi^2}  \le[
        \frac{2}{i\w} \frac{A_z^{(2)}(\w)}{A_z^{(0)}(\w)} + \frac{i \w}{2} \ri]
    \label{eq:sigma_zz_formula}
\end{align}\end{subequations}
Since \(\s_{xx}\) and \(\s_{yy}\) are equal, we will show results only for \(\s_{xx}\). The derivation of equation~\eqref{eq:sigma_formula} may be found in appendix~\ref{app:conductivities}.

To summarise, the conductivities are obtained by solving the equations of motion for \(A_n(\w;r)\) in equations~\eqref{eq:A_z_eom} and~\eqref{eq:A_plus_minus_eom} subject to the ingoing boundary conditions in equation~\eqref{eq:ingoing_boundary_conditions}. From the behaviour of the resulting solutions at \(r=0\) we then read off \(A_n^{(0)}(\w)\) and \(A_n^{(2)}(\w)\), which we substitute into equation~\eqref{eq:sigma_formula}. Usually  this procedure must be carried out numerically. However, as we discuss in the next subsection, for massless hypermultiplets \(m=0\) the conductivities may be obtained analytically.

Reality of \(A_i(t,r)\) requires that the Fourier modes obey the condition \(A_\pm(\w;r) = A_\mp(-\w;r)^*\) and \(A_z(\w;r) = A_z(-\w;r)^*\). This then implies that \(A_\pm^{(0,2)}(\w) = A_\mp^{(0,2)}(-\w)^*\) and \(A_z^{(0,2)}(\w) = A_z^{(0,2)}(-\w)^*\). These conditions combined with equation~\eqref{eq:sigma_formula} make manifest that the real parts of the conductivities are even functions of \(\w\), while the imaginary parts are odd functions of \(\w\). This is true even though non-zero \(b\) breaks time-reversal symmetry. For example, we may rewrite the real and imaginary parts of \(\s_{xx}\) as
\begin{equation}\begin{aligned} \label{eq:sigma_xx_real_im}
    \Re \s_{xx}(\w) 
    &= \frac{N_f N_c}{4 \pi^2} \frac{1}{\w} \Im \le(\frac{A_+^{(2)}(\w)}{A_+^{(0)}(\w)} - \frac{A_+^{(2)}(-\w)}{A_+^{(0)}(-\w)} \ri),
    \\
    \Im \s_{xx}(\w) &= \frac{N_f N_c}{4\pi^2} \le[- \frac{1}{\w} \Re \le(\frac{A_+^{(2)}(\w)}{A_+^{(0)}(\w)} + \frac{A_+^{(2)}(-\w)}{A_+^{(0)}(-\w)}  \ri) + \frac{\w}{2}\ri],
\end{aligned}\end{equation}
which are manifestly even and odd functions of \(\w\), respectively. Similar results hold for \(\s_{xy}\) and \(\s_{zz}\). As a consequence, we will only show results for the conductivities at positive frequency.

The conductivities we present will be functions of four parameters, each with units of mass: \(m\), \(T\), \(b\), and \(\w\). The conductivities themselves also have units of mass. In what follows, we will mostly measure dimensionful quantities in units of \(b\), plotting \(\s/b\) as a function of \(T/b\), \(\w/b\), and \(m/b\sqrt{\l}\), the factor of \(\sqrt{\l}\) in the latter arising naturally from holography. The exception is when we present results at \(m=0\), in which case the \(b\)-dependence of the conductivities drops out and we instead present results in units of \(T\), and some occasions where we present the conductivities in units of their zero mass limits.

\subsection{Conductivities of massless hypermultiplets}

For \(m=0\) the embedding equation~\eqref{eq:embedding_equation} is solved by \(R(r) = 0\). This greatly simplifies the equations of motion for the gauge field fluctuations; for \(R(r)=0\) equations~\eqref{eq:A_z_eom} and~\eqref{eq:A_plus_minus_eom} both become of the same form,
\begin{equation} \label{eq:A_plus_minus_eom_m0}
    (r^4 - \r_h^4) \le(\frac{r^4 - \r_h^4}{r} A_n' \ri)' + \frac{\w^2}{r} (r^4 + \r_h^4) A_n = 0.
\end{equation}
Since the equations of motion for \(A_+\) and \(A_-\) are identical at \(m=0\), solutions obeying the same ingoing boundary conditions will have \(A_+^{(0)}(\w) = A_-^{(0)}(\w)\) and \(A_+^{(2)}(\w) = A_-^{(2)}(\w)\). The \(A_+^{(2)}/A_+^{(0)}\) and \(A_-^{(2)}/A_-^{(0)}\) terms cancel in the expression for \(\s_{xy}\) in equation~\eqref{eq:sigma_xy_formula}, leading to an AC Hall conductivity that is independent of both frequency and temperature, equal to the DC conductivity found in ref.~\cite{BitaghsirFadafan:2020lkh},
\begin{equation} \label{eq:sigma_xy_T0}
    \le.\s_{xy}(\w)\ri|_{m=0} = \frac{N_f N_c}{4 \pi^2} b.
\end{equation}
Further, since the equation of motion for \(A_z\) takes the same form as those for \(A_\pm\) the longitudinal conductivities are equal at \(m=0\), \(\le.\s_{zz}(\w)\ri|_{m=0} = \le.\s_{xx}(\w)\ri|_{m=0}\).

Since equation~\eqref{eq:A_plus_minus_eom_m0} is independent of \(b\), its solutions are the same as for \(b=0\). For \(T=0\) (corresponding to \(\r_h =0 \)) the solution obeying ingoing boundary conditions is
\begin{equation} \label{eq:zero_T_massless_gauge_field}
    A_n(\w;r) = e^{3 \pi i/4} \sqrt{\frac{\pi \w}{2}} \frac{H_1^{(1)}(\w/r)}{r},
\end{equation}
where \(H_\a^{(1)}\) is a Hankel function of the first kind. Expanding this solution at large \(r\) we read off \(A_n^{(0)}(\w)\) and \(A_n^{(2)}(\w)\), which we then substitute into equations~\eqref{eq:sigma_xx_formula} and~\eqref{eq:sigma_zz_formula} to obtain the longitudinal conductivities at \(m = T = 0\),
\begin{equation} \label{eq:sigma_xx_m0_T0}
    \le.\s_{xx}(\w)\ri|_{m=T=0}  = \le.\s_{zz}(\w)\ri|_{m=T=0}  =  \frac{N_f N_c}{4 \pi^2} \le[\frac{\pi}{2} + i \le( \g -  \log 2\ri) \ri] \omega,
\end{equation}
where \(\g \approx 0.577\) is the Euler--Mascheroni constant.

\begin{figure}
    \begin{center}
        \includegraphics{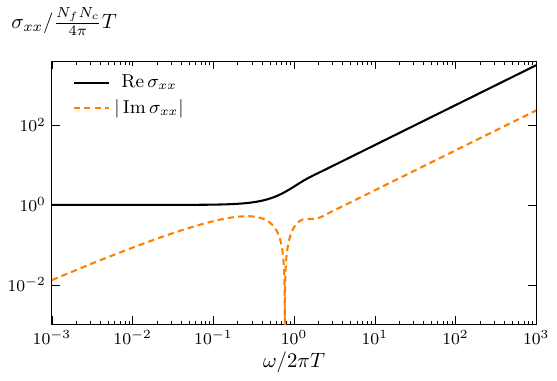}
    \end{center}
    \caption{Logarithmic plot of the longitudinal AC conductivity at \(m=0\) and non-zero temperature, written in equation~\eqref{eq:sigma_xx_m0}, as a function of frequency \(\w\). The imaginary part is positive at low frequencies and negative at large frequencies. For \(\w/T\gg1\) we find that \(\s_{xx}\) becomes linear in frequency, asymptoting to the zero temperature result in equation~\eqref{eq:sigma_xx_m0_T0}.}
    \label{fig:sigma_xx_m0}
\end{figure}

The solution to equation~\eqref{eq:A_plus_minus_eom_m0} obeying ingoing boundary conditions at non-zero temperature was found in ref.~\cite{Myers:2007we}. In our coordinate system it is
\begin{equation} \label{eq:m0_exact_sol}
    A_n(\w;r) = \le(\frac{r^2 - \r_h^2}{r^2 + \r_h^2}\ri)^{-i\fw} {}_2 F_1 \le( \frac{1 - i}{4} \fw,- \frac{1+i}{2} \fw;1 - i\fw; \le( \frac{r^2 - \r_h^2}{r^2 + \r_h^2}\ri)^2 \ri),
\end{equation}
where we have introduced the short-hand notation \(\fw = \w/2\pi T\). Again expanding this solution at large \(r\) to find \(A_n^{(0)}(\w)\) and \(A_n^{(2)}(\w)\), we obtain the conductivities~\cite{Myers:2007we}
\begin{equation}\begin{aligned} \label{eq:sigma_xx_m0}
    \le.\s_{xx}(\w)\ri|_{m=0} = \frac{N_f N_c}{4} T \biggl[
         \fw  \frac{\sinh(\pi\fw) + i  \sin(\pi\fw)}{\cosh(\pi \fw) - \cos(\pi \fw)}
        + & \frac{2i\fw}{\pi} \Re \y\le( \frac{1+i}{2} \fw \ri)
        \\
        &+ \frac{i}{\pi}
        + \frac{i\fw}{\pi} \le(2\g  - \log(2 \fw^2) \ri)
    \biggr],
\end{aligned}\end{equation}
where \(\y\) is the digamma function. This conductivity is plotted in figure~\ref{fig:sigma_xx_m0}. It interpolates between the \(m=0\) value of the DC conductivity found in ref.~\cite{BitaghsirFadafan:2020lkh}, \(\le.\s_{xx}(0) \ri|_{m=0} = \frac{N_f N_c}{4 \pi}T\), at low frequencies and the zero temperature result in equation~\eqref{eq:sigma_xx_m0_T0} for \(\w \gg T\).

\subsection{Conductivities of massive hypermultiplets}

For \(m \neq 0\) we must solve both the embedding equation~\eqref{eq:embedding_equation} for \(R(r)\) and equation~\eqref{eq:A_plus_minus_eom} for \(A_n(\w;r)\) numerically. We have done so using two methods. The first is a shooting method, which works as follows. We use the IR asymptotics for \(R(r)\) written in equation~\eqref{eq:bh_embedding_ir} and~\eqref{eq:bh_embedding_ir_T0} for some value of \(r_h\) or \(\h\), to set boundary conditions at small \(r\), and integrate to large \(r\) using \texttt{NDSolve} in Mathematica.  The value of \(M\) is obtained from the \(r \to \infty \) limit of the solution. We then tune the IR coefficient \(r_h\) or \(\h\) to obtain the desired value of \(M\). Then, we substitute the solution for \(R(r)\) into the gauge field equations~\eqref{eq:A_z_eom} and~\eqref{eq:A_plus_minus_eom} and solve these for \(A_n(\w;r)\) with \texttt{NDSolve}, using equation~\eqref{eq:ingoing_boundary_conditions} to set ingoing boundary conditions at small \(r\). From the behaviour of these solutions near \(r \to \infty\) we determine \(A_n^{(0)}(\w)\) and \(A_n^{(2)}(\w)\), which we then substitute into equation~\eqref{eq:sigma_formula} to obtain the conductivities.

The other numerical method that we used is a pseudospectral collocation method, detailed in appendix~\ref{app:pseudospectral}. The advantage of this method is that with it we can simultaneously impose boundary conditions in the IR and UV, eliminating the need to tune the IR coefficient \(r_h\) or \(\h\). The two numerical methods agree where we have checked them against each other. All of the plots we show of the conductivities use results obtained from the pseudospectral method, which we found was faster and converged better at large frequencies.

\begin{figure}[!htbp]
    \begin{center}\includegraphics{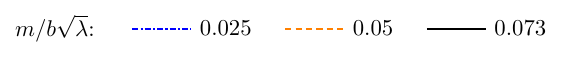}\end{center}\vspace{-12pt}
    \begin{subfigure}{\textwidth}
        \includegraphics{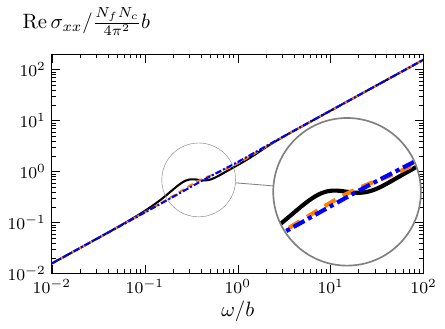}
        \includegraphics{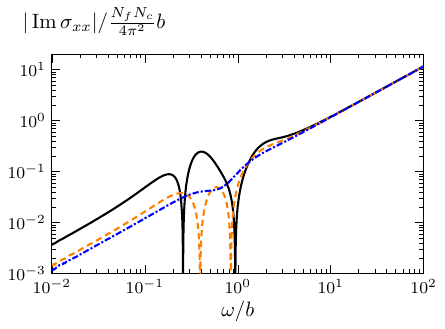}
    \end{subfigure}
    \begin{subfigure}{\textwidth}
        \includegraphics{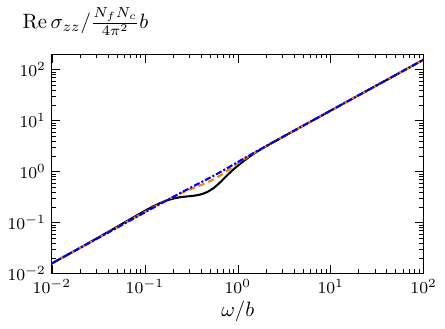}
        \includegraphics{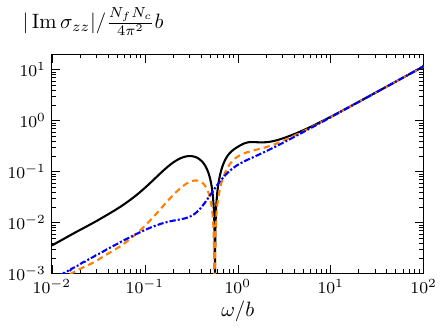}
    \end{subfigure}
    \begin{subfigure}{\textwidth}
        \includegraphics{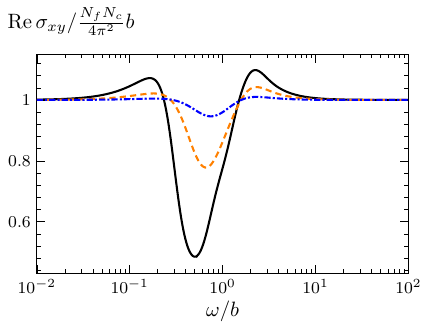}
        \includegraphics{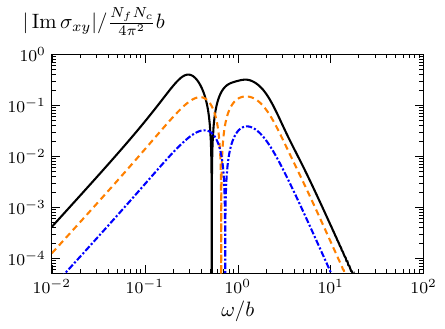}
    \end{subfigure}
    \caption{The AC conductivities as functions of frequency, at zero temperature and for sample values of the hypermultiplet masses. The black curves show \(m = 0.073\, b\sqrt{\l}\), just below the transition to the insulating phase (the phase transition is at \(m \approx 0.0733 \, b\sqrt{\l}\)). We plot the absolute values of the imaginary values to be able to show them on logarithmic axes. The signs can be inferred from the information that \(\Im \s_{xx}\), \(\Im \s_{xy}\) and \(\Im \s_{zz}\) are all negative at large frequencies.}
    \label{fig:T0_conductivities}
\end{figure}

\subsubsection{Zero temperature}

In figure~\ref{fig:T0_conductivities} we show numerical results for the AC conductivities at \(T=0\) and three different values of \(m/b\sqrt{\l}\). For the smallest mass plotted \(m = 0.025 \, b\sqrt{\l}\), (dot-dashed blue curves in the figure), the longitudinal conductivities behave qualitatively similarly to the zero mass result plotted in figure~\ref{fig:sigma_xx_m0}, and the Hall conductivity is almost constant, similar to the exact result at \(T=0\) in equation~\eqref{eq:sigma_xy_T0}.

As the mass is raised, first to \(m = 0.05 \, b\sqrt{\l}\) (dashed orange) and then to \(m = 0.073 \, b \sqrt{\l}\) (solid black), features start to develop in the conductivities. The real part of the longitudinal conductivity \(\s_{xx}\) orthogonal to the applied axial field develops a bump, while the real parts of \(\s_{zz}\) and the Hall conductivity \(\s_{xy}\) develop dips/troughs. These features occur at frequencies of order slightly less than \(b\); for example at \(m = 0.073 \, b \sqrt{\l}\) the local maximum in \(\Re \s_{xx}\) occurs at \(\w \approx 0.26 \, b\), while the local minimum in \(\Re \s_{xy}\) occurs at \(\w \approx 0.50 \, b\).

\begin{figure}[!t]
        \includegraphics{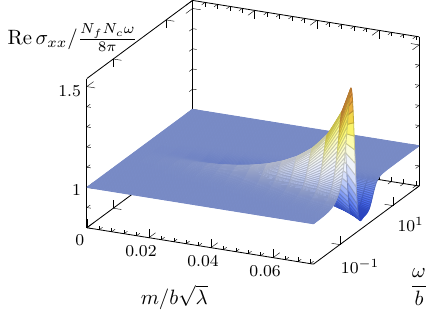}
        \includegraphics{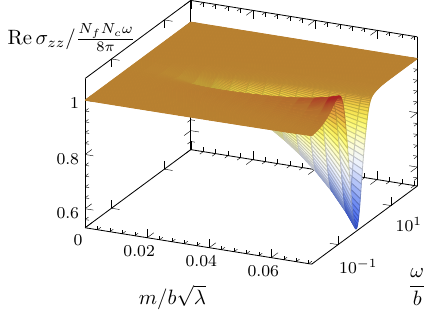}
        \begin{center}
            \includegraphics{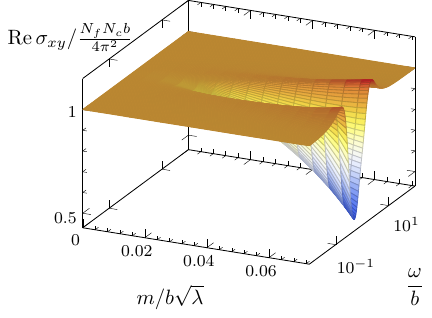}
        \end{center}
    \caption{Real parts of the AC conductivities at zero temperature as functions of mass and frequency, normalised to their values at \(m=0\). These plots show how the peaks and troughs in the conductivities develop as the mass is increased towards the phase transition at \(m \approx 0.073 \, b \sqrt{\l}\)}
    \label{fig:conductivity_T0_3d}
\end{figure}

The largest mass for which we plot the conductivities in figure~\ref{fig:T0_conductivities} is \(m= 0.073 \, b \sqrt{\l}\), which is just below that of the phase transition to the insulating phase. In figure~\ref{fig:conductivity_T0_3d} we show three-dimensional plots of the real parts of the conductivities in the \((\w/b,m/b\sqrt{\l})\) plane, showing how the height of the peak and the depths of the troughs increase as we increase \(m/b\sqrt{\l}\) towards the phase transition. In figure~\ref{fig:conductivity_T0_3d} we have normalised the conductivities to their values at \(m=0\), which for \(\s_{xx}\) and \(\s_{zz}\) makes the features that develop near the phase transition more visible against the approximately linear growth of the conductivities with frequency.

Features visible in both figures~\ref{fig:T0_conductivities} and~\ref{fig:conductivity_T0_3d} are that the conductivities become independent of mass for \(\w/b \gg 1\), and that the real parts of the conductivities are independent of mass for \(\w/b \ll 1\). Moreover, the longitudinal conductivities are linear in frequency with the same slope in both limits. These behaviours arise due to the scaling behaviour of the black hole embeddings in the UV and the IR. From equation~\eqref{eq:induced_metric}, we see that in the the far UV \(r \to \infty\) the induced metric on the D7-branes' world-volume is that of \(\ads[5] \times \sph[3]\) with both \ads[5] and \sph[3] having radius of curvature \(L\), independent of the hypermultiplet mass. The large frequency behaviour of the conductivities is controlled by the UV behaviour of the geometry, and hence is also independent of the hypermultiplet mass. Similarly, for zero-temperature black hole embeddings, the IR \(r \to 0\) limit of the induced metric is also \(\ads[5] \times \sph[3]\) with curvature radius \(L\). The low-frequency limits of the imaginary parts of holographic Green's functions---and therefore through the Kubo formula~\eqref{eq:conductivity_kubo_formula} the real parts of the conductivities---are determined by the IR geometry~\cite{Cubrovic:2009ye,Faulkner:2009wj,Faulkner:2011tm}.

The fact that the DC limit of \(\s_{xy}\) is independent of mass in the WSM phase was established in ref.~\cite{BitaghsirFadafan:2020lkh}. One way to derive the low-frequency behaviour of the real parts of \(\s_{xx}\) and \(\s_{zz}\) is as follows~\cite{Gubser:2008wz}. Using the equations of motion~\eqref{eq:A_z_eom} and~\eqref{eq:A_plus_minus_eom}, it is straightforward to show that the following quantities, which we shall refer to as fluxes, are independent of \(r\),
\begin{align} \label{eq:fluxes}
    \cF_+(\w)\!&=\!\frac{r^3 \sqrt{(r^2+R^2)^2 + b^2 R^2}}{2i\w(r^2 + R^2) \sqrt{1 + R'^2}}\le[A_+(\w;r) \p_r A_-(-\w;r) - A_-(-\w;r) \p_r A_+(\w;r)\ri],
    \\
    \cF_z(\w)\!&=\!\frac{r^3  (r^2 + R^2) }{2i\w\sqrt{(r^2 + R^2)^2 + b^2 R^2}\sqrt{(1+R'^2)}} \!\le[ A_z(\w;r) \p_r A_z(-\w;r) - A_z(-\w;r) \p_r A_z(\w;r) \ri],
    \nonumber
\end{align}
Evaluating these fluxes near the boundary, one finds
\begin{equation} \label{eq:fluxes_near_boundary}
    \cF_+(\w) = \frac{2}{\w} |A_+^{(0)}(\w)|^2 \Im \le(\frac{A_+^{(2)}(\w)}{A_+^{(0)}(\w)}\ri),
    \qquad
    \cF_z(\w) = \frac{2}{\w} |A_z^{(0)}(\w)|^2 \Im \le(\frac{A_z^{(2)}(\w)}{A_z^{(0)}(\w)}\ri),
\end{equation}
where we have made use of the fact that reality of \(A_x(t,r)\) and \(A_y(t,r)\) requires that \(A_+(\w;r) = A_-(-\w;r)^*\), while reality of \(A_z(t,r)\) requires \(A_z(\w;r) = A_z(-\w;r)^*\).

Using equations~\eqref{eq:sigma_formula} and~\eqref{eq:fluxes_near_boundary}, we see that we may rewrite the real parts of the longitudinal conductivities in terms of the fluxes
\begin{equation}\label{eq:sigma_xx_zz_flux}
    \Re \s_{xx}(\w) = \frac{N_f N_c}{8 \pi^2}  \le(\frac{\cF_+(\w)}{|A_+^{(0)}(\w)|^2} +  \frac{\cF_+(-\w)}{|A_+^{(0)}(-\w)|^2}\ri),
    \qquad
    \Re \s_{zz}(\w) = \frac{N_f N_c}{4 \pi^2 } \frac{ \cF_z(\w)}{|A_z^{(0)}(\w)|^2}.
\end{equation}
Evaluating the fluxes at small \(r\), using the ingoing boundary conditions~\eqref{eq:ingoing_boundary_conditions} and that for black hole embeddings we have \(R \ll r\) as \(r \to 0\), one finds
\begin{equation}
    \cF_+(\w) = \cF_z(\w) = 1.
\end{equation}
Substituting this into equation~\eqref{eq:sigma_xx_zz_flux}, we obtain the formulas
\begin{equation}\label{eq:re_sigma_xx_zz_formula}
    \Re \s_{xx}(\w) = \frac{N_f N_c}{8 \pi^2}  \le(\frac{1}{|A_+^{(0)}(\w)|^2} +  \frac{1}{|A_+^{(0)}(-\w)|^2}\ri),
    \qquad
    \Re \s_{zz}(\w) = \frac{N_f N_c}{4 \pi^2 } \frac{1}{|A_z^{(0)}(\w)|^2}.
\end{equation}
So far no approximation has been made. We will now make a small frequency approximation, solving the equations of motion for the fluctuations at \(\w \ll b\) in order to obtain \(A_n^{(0)}(\w)\), and therefore determine \(\Re \s_{xx}\) and \(\Re \s_{zz}\) from equation~\eqref{eq:re_sigma_xx_zz_formula}.

At leading order at small non-zero \(\w\), one would be tempted to set \(\w=0\) in the equations of motion~\eqref{eq:A_z_eom} and~\eqref{eq:A_plus_minus_eom} for \(A_n\). However, this does not work because the coefficient of \(\w^2\) in these equations diverges at \(r \to 0\). Instead, the correct approach is to solve for the gauge field fluctuations using a matching method~\cite{Cubrovic:2009ye,Faulkner:2009wj,Faulkner:2011tm}; one solves the equations of motion for non-zero frequency close to \(r=0\) with the approximation \(R(r) \approx 0\) (since \(R \ll r\) for the black hole emebddings as \(r \to 0\)) and for zero frequency away from \(r=0\), matching these two solutions in the region where both approximations are valid. The solution near \(r=0\) is the zero mass solution written in equation~\eqref{eq:zero_T_massless_gauge_field}, while the zero frequency solution is \(A_n(\w;r) = A_n^{(0)}(\w)\) constant. Matching these two solutions, one finds
\begin{equation}
    A_n^{(0)}(\w) = \lim_{\w/r \to 0} e^{3 \pi i/4} \sqrt{\frac{\pi \w}{2}} \frac{H_1^{(1)}(\w/r)}{r}  = \mathrm{sign}(\w) e^{i\pi/4} \sqrt{\frac{2}{\pi \w}}.
\end{equation}
Substituting this into equation~\eqref{eq:re_sigma_xx_zz_formula}, we obtain the results
\begin{equation}
    \Re \s_{xx}(\w) = \Re \s_{zz}(\w) = \frac{N_f N_c}{8\pi}\w,
    \qquad
    \w \ll b,
\end{equation}
matching the real parts of \(\s_{xx}\) and \(\s_{zz}\) at \(m=0\) written in equation~\eqref{eq:sigma_xx_m0_T0}, as anticipated from the numerical results.

The imaginary part of \(\s_{xy}\) in equation~\eqref{eq:sigma_xy_formula} may also be written in terms of \(\cF_+\),
\begin{equation}
    \Im \s_{xy}(\w) = \frac{N_f N_c}{8\pi^2} \le(\frac{\cF_+(-\w)}{|A_+(-\w)|^2} - \frac{\cF_+(\w)}{|A_+(\w)|^2} \ri).
\end{equation}
The analysis of the preceding paragraphs then implies that \(\Im \s_{xy} \to 0\) as \(\w \to 0\), but does not tell us the rate that at which \(\Im \s_{xy}\) vanishes. Numerically we find \(\Im \s_{xy} \propto \w^2/b\) at small \(\w/b\), with a coefficient that (as can be seen in figure~\ref{fig:T0_conductivities}) depends on \(m/b\sqrt{\l}\).

\subsubsection{Non-zero temperature}

We now turn to the conductivities at both non-zero temperature and non-zero hypermultiplet mass. First, we compute the DC conductivity \(\s_{zz}(\w=0)\). Unlike \(\s_{xx}\) and \(\s_{xy}\), the DC value of \(\s_{zz}\) for the D3/D7 WSM was not computed in ref.~\cite{BitaghsirFadafan:2020lkh} since it was not amenable to the Karch--O'Bannon method used there. However there is no problem computing \(\s_{zz}(\w=0)\) using the appropriate limit of the Kubo formula.

One way to obtain the DC conductivity numerically would be to take the zero frequency limit in our numerical computation of the AC conductivities. However, it is also possible to derive a formula for the DC conductivity that depends only on the properties of the embedding \(R(r)\). We will use the latter approach, since it allows for a more direct determination of the DC conductivity.

To derive this formula, one could use the same method of fluxes that we used to obtain the low-frequency behaviour of \(\Re \s_{zz}(\w)\) at zero temperature. However a more transparent method is the membrane paradigm approach of ref.~\cite{Iqbal:2008by}, which works as follows. Using the holographic renormalisation results in appendix~\ref{app:conductivities}, one finds that the radial canonical momentum conjugate to \(A_z\) (i.e. the derivative of the Lagrangian density with respect to \(\p_r A_z\)) is
\begin{equation}
    \Pi_z(\w;r)  = - \frac{N_f N_c}{4\pi^2} \le[
        \frac{\r^2 r^3 \, f \sqrt{h}}{\sqrt{(\r^4 h + b^2 R^2)(1 + R'^2)}} \p_r A_z(\w;r) + \frac{\w^2}{2} A_z(\w;r) \log(r^2/\w^2) 
    \ri],
\end{equation}
where the term containing \(\log(r^2/\w^2)\) arises from a counterterm. Expanding for large \(r\), one finds
\begin{equation}
    \Pi_z(\w;r) = \frac{N_f N_c}{4\pi^2} \le(2 A_z^{(2)}(\w) - \frac{\w^2}{2} A_z^{(0)}(\w) \ri) + O(r^{-1}).
\end{equation}
Comparing to the formula for \(\s_{zz}\) in equation~\eqref{eq:sigma_zz_formula}, we find that we may write
\begin{equation} \label{eq:sigma_zz_from_momentum}
    \s_{zz}(\w) = \lim_{r \to \infty} \frac{\Pi_z(\w;r)}{i \w A_z(\w;r)} = \frac{N_f N_c}{4\pi^2} \lim_{r\to \infty} \le[ P_z(\w;r)  - \frac{\w^2}{2} \log(r^2/\w^2)\ri],
\end{equation}
where on the right-hand side we have defined
\begin{equation}\begin{aligned}
    P_z(\w;r)
    &\equiv \frac{4 \pi^2}{N_f N_c} \le(\frac{\Pi_z(\w;r)}{i \w A_z(\w;r)} + \frac{\w^2}{2} \log(r^2/\w^2)\ri),
    \\
    &=-\frac{\r^2 r^3 \, f \sqrt{h}}{\sqrt{(\r^4 h + b^2 R^2)(1 + R'^2)}} \frac{\p_r A_z(\w;r)}{A_z(\w;r)}.
\end{aligned}\end{equation}

The key observation of ref.~\cite{Iqbal:2008by} is that in the limit \(\w \to 0\), \(P_z(\w;r)\) becomes independent of \(r\) and the logarithmic term drops from equation~\eqref{eq:sigma_zz_from_momentum}, so that we may obtain the DC value of \(\s_{zz}\) by evaluating \(P_z(\w;r)\) at the horizon. Indeed, using the equation of motion~\eqref{eq:A_z_eom}, one finds that \(P_z(\w;r)\) obeys the equation of motion
\begin{equation}
    \frac{\r^2 r^3 \, f \sqrt{h}}{\sqrt{(\r^4 h + b^2 R^2)(1 + R'^2)}} \p_r P_z(\w;r) = i \w \le(\frac{r^6 h}{\r^4 h + b^2 R^2} - P_z(\w;r)^2 \ri).
\end{equation}
We see explicitly that \(\p_r P_z(\w;r) = O(\w)\), and therefore to leading order at zero frequency, \(P_z(\w;r)\) evaluated at the boundary is the same as \(P_z(\w;r)\) evaluated at the black hole horizon. We can therefore write the DC limit of equation~\eqref{eq:sigma_zz_from_momentum} as
\begin{equation} \label{eq:sigma_zz_dc}
    \s_{zz}(\w = 0) = \frac{N_f N_c}{4\pi^2} \lim_{\w \to 0} P_z(\w;r_h) = \frac{N_f N_c}{2\pi^2} \frac{r_h^3}{\sqrt{2 \r_h^4 + b^2 (\r_h^2 - r_h^2)}},
\end{equation}
Where the right-hand side is evaluated by substituting the ingoing boundary conditions for \(A_z(\w;r)\) in equation~\eqref{eq:ingoing_boundary_conditions} and the near-horizon behaviour of \(R(r)\) written in equation~\eqref{eq:bh_embedding_ir}. Note that we must still use numerics to determine how \(\s_{zz}(\w=0)\) depends on \(m/b\sqrt{\l}\), because the only way to find out what value of \(r_h\) corresponds to what value of \(m/b\sqrt{\l}\) is to numerically solve the equation of motion for \(R(r)\).

\begin{figure}[!tp]
    \begin{center}
    \includegraphics{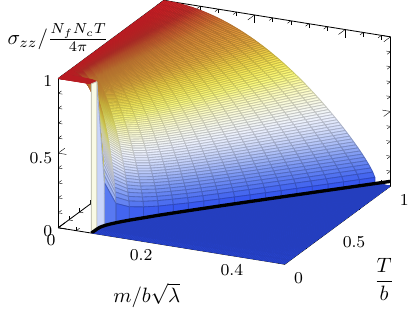}
    \hspace{1cm}
    \includegraphics{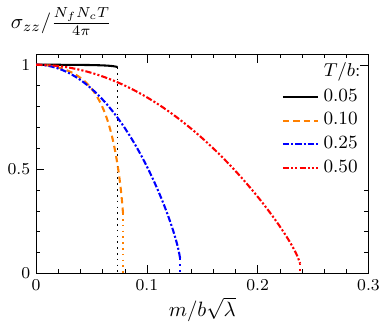}
    \end{center}
    \caption{Longitudinal DC conductivity parallel to the applied axial field, \(\s_{zz}(\w=0)\), plotted normalised to its value for \(m=0\), computed from equation~\eqref{eq:sigma_zz_dc}. The left-hand panel shows a three-dimensional plot of \(\s_{zz}(\w=0)\)  as a function of dimensionless mass \(m/b\sqrt{\l}\) and temperature \(T/b\). The right-hand panel shows several slices through three-dimensional plot, at various fixed temperatures. The DC conductivity is discontinuous, dropping to zero at the transition to the insulating phase, the location of which is indicated with the thick black curve in the left-hand panel.}
    \label{fig:sigma_zz_dc}
\end{figure}

Our numerical results for \(\s_{zz}\) at \(\w=0\) as a function of \(T/b\) and \(m/b\sqrt{\l}\), obtained using equation~\eqref{eq:sigma_zz_dc}, are plotted in figure~\ref{fig:sigma_zz_dc}. We normalise the DC conductivity to its massless value, corresponding to \(r_h=\r_h\) in equation~\eqref{eq:sigma_zz_dc},
\begin{equation}
        \le.\s_{zz}(\w = 0)\ri|_{m=0} = \frac{N_f N_c}{4\pi^2} \lim_{\w \to 0} P_z(\w;r_h) = \frac{N_f N_c}{4\pi} T.
\end{equation}
When normalised in this way, the functional form of \(\s_{zz}(\w=0)\) is almost identical to that of \(\s_{xy}(\w=0)\) normalised to its \(m=0\) value, plotted in figure~\ref{fig:old_dc_conductivities}. We find that \(\s_{zz}/T\) is approximately a step function of \(m/b\sqrt{\l}\) at low temperatures; approximately constant in the WSM phase and then dropping to zero at the transition to the insulating phase. At larger temperatures, \(\s_{zz}/T\) decreases with increasing \(m/b\sqrt{\l}\) across the WSM phase, reducing the size of the discontinuous drop at the phase transition.

\begin{figure}[!htbp]
    \begin{center}\includegraphics{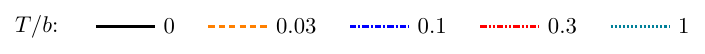}\end{center}\vspace{-16pt}
    \begin{subfigure}{\textwidth}
        \includegraphics{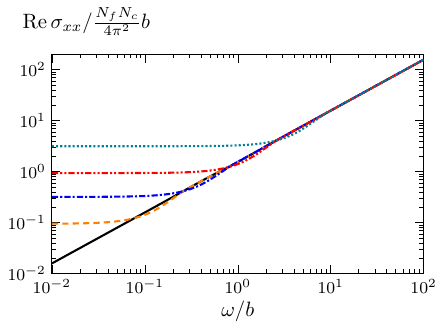}
        \includegraphics{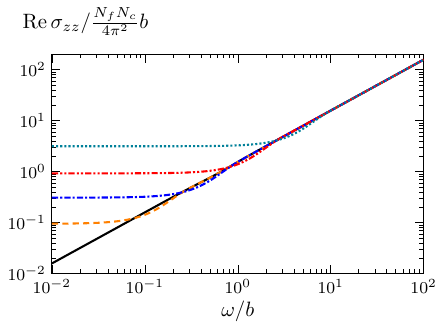}
        \vspace{-20pt}
        \caption{\(m = 0.025 \,  b \sqrt{\l}\)}
    \end{subfigure}
    \begin{subfigure}{\textwidth}
        \includegraphics{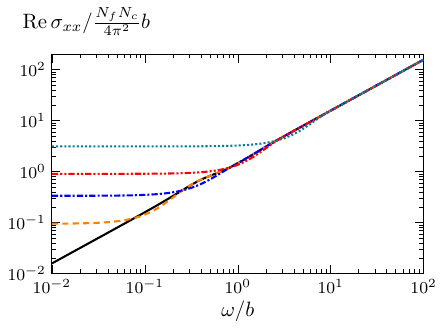}
        \includegraphics{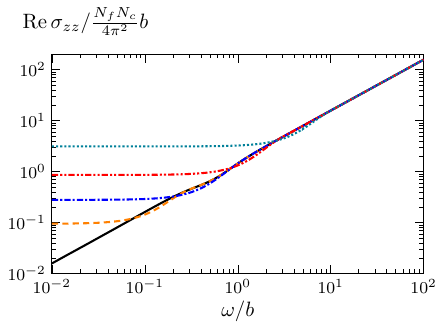}
        \vspace{-20pt}
        \caption{\(m = 0.05 \, b \sqrt{\l}\)}
    \end{subfigure}
    \begin{subfigure}{\textwidth}
        \includegraphics{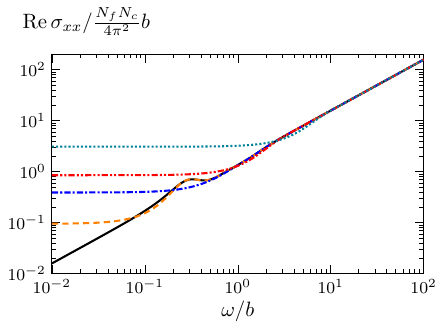}
        \includegraphics{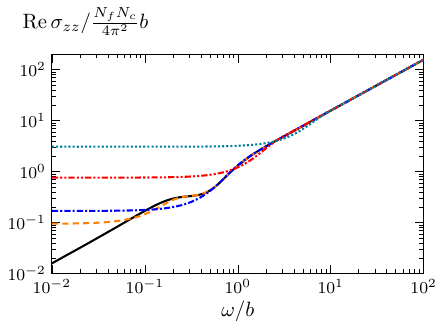}
        \vspace{-20pt}
        \caption{\(m = 0.073 \, b \sqrt{\l}\)}
        \label{fig:conductivity_real_0p073}
    \end{subfigure}
    \caption{Real parts of the longitudinal AC conductivities as functions of frequency, at various sample masses \(m/b\sqrt{\l}\) and temperatures \(T/b\). The top row shows a relatively small value of the mass, while the bottom row shows a mass very close to the location of the phase transition at \(T=0\).
    }
    \label{fig:conductivity_real}
\end{figure}

We now move on to the AC conductivities. In figure~\ref{fig:conductivity_real} we plot the real parts of \(\s_{xx}\) and \(\s_{zz}\), at various sample values of \(T/b\) and \(m/b\sqrt{\l}\). For reference, we also show the zero temperature results. In all cases, the results are qualitatively the same, and also similar to the result at \(m=0\) plotted in figure~\ref{fig:sigma_xx_m0}. The real parts of the conductivities interpolate between their DC values at small \(\w/b\), and the massless, zero temperature result written in equation~\eqref{eq:sigma_xx_m0_T0} at large frequencies. As previously commented on, at zero temperature the conductivities show more features (peaks and troughs) close to the phase transition at \(m \approx 0.0733 \, b\sqrt{\l}\). As can be seen in figure~\ref{fig:conductivity_real_0p073}, these features are rapidly washed out as the temperature is raised from zero, having more-or-less disappeared by \(T  =0.1 \, b\). The imaginary parts of the conductivities plotted in figure~\ref{fig:conductivity_real} are plotted in figure~\ref{fig:conductivity_imaginary} in appendix~\ref{app:plots}.

\begin{figure}[!htbp]
    \begin{center}\includegraphics{figs/conductivity_legend.pdf}\end{center}\vspace{-12pt}
    \begin{subfigure}{0.5\textwidth}
        \includegraphics{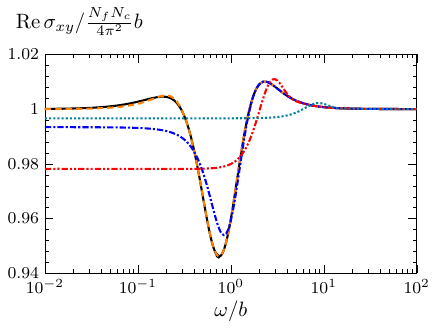}
        \caption{\(m = 0.025 \,  b \sqrt{\l}\)}
    \end{subfigure}\begin{subfigure}{0.5\textwidth}
        \includegraphics{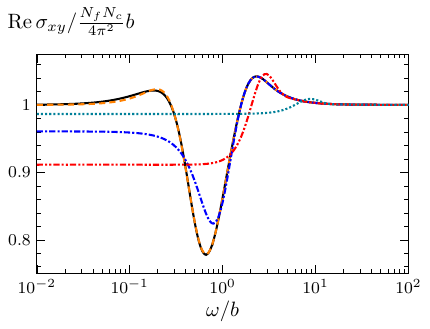}
        \caption{\(m = 0.05 \, b \sqrt{\l}\)}
    \end{subfigure}
    \begin{center}\begin{subfigure}{0.5\textwidth}
        \includegraphics{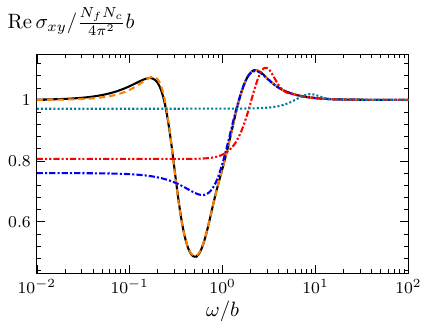}
        \caption{\(m = 0.073 \, b \sqrt{\l}\)}
    \end{subfigure}\end{center}
    \caption{Real part of the AC Hall conductivity as a function of frequency, at various sample masses \(m/b\sqrt{\l}\) and temperatures \(T/b\). As the temperature is raised, the trough in the Hall conductivity disappears.}
    \label{fig:hall_conductivity_real}
\end{figure}

In figure~\ref{fig:hall_conductivity_real} we plot the real part of the Hall conductivity \(\Re \s_{xy}\) for the same temperature and masses as we plotted \(\Re \s_{xx}\) and \(\Re \s_{zz}\). The story for the Hall conductivity is similar to that for the longitudinal conductivities. At low temperatures \(\Re \s_{xy}\) exhibits a trough that grows deeper as the mass is raised towards the transition to the insulating phase. As the temperature is raised this trough disappears.  The imaginary parts of the conductivities plotted in figure~\ref{fig:hall_conductivity_real} are plotted in figure~\ref{fig:hall_conductivity_imaginary} in appendix~\ref{app:plots}.

\section{Quasinormal modes}
\label{sec:qnm}

As seen in the previous section, at low or zero temperature the AC conductivities exhibit peaks and troughs as functions of frequency. These features, particularly the peak in \(\s_{xx}\), might be expected to arise due to the presence of poles in the retarded Green's functions in the complex frequency plane that are close to the real axis at low temperatures and for masses close to the phase transition. In this section we investigate this intuition, finding that indeed there are poles close to the real axis, and arguing that their presence is related to the first-order nature of the phase transition.

Poles in retarded Green's functions are holographically dual to the frequencies of quasinormal modes~\cite{Birmingham:2001pj,Son:2002sd,Kovtun:2005ev}, the complex frequency eigenvalues for which one can have perturbations of the gravitational system that are ingoing at the black hole horizon and normalisable at the conformal boundary. The quasinormal modes of the pseudoscalar \(\f(r)\) in the flavour brane WSM, dual to poles of the two-point function of a pseudoscalar operator in the dual QFT, were studied in ref.~\cite{Atashi:2022ufl}.

In our context, the quasinormal modes dual to poles in the retarded Green's functions of the charge current---and therefore dual to poles in the conductivities---are the values of \(\w\) for which the gauge field fluctuation equations~\eqref{eq:A_z_eom} and~\eqref{eq:A_plus_minus_eom} admit solutions satisfying ingoing boundary conditions~\eqref{eq:ingoing_boundary_conditions} in the IR and that are normalisable in the UV, \(A_n^{(0)}(\w) = 0\). The quasinormal modes in the \(A_\pm\) channels are dual to poles in \(\s_{xx}\) and \(\s_{xy}\), since from equation~\eqref{eq:sigma_formula} we see that when \(A_+^{(0)}(\w)=0\) or \(A_-^{(0)}(\w)=0\), both \(\s_{xx}\) and \(\s_{xy}\) will diverge. Similarly, the quasinormal modes in the \(A_z\) channel are dual to poles in \(\s_{zz}\), since when \(A_z^{(0)}(\w)\) vanishes, then according to equation~\eqref{eq:sigma_formula}, \(\s_{zz}\) will diverge.

For \(m=0\) and non-zero temperature, the quasinormal modes may be obtained analytically, from the near-boundary asymptotics of the solution for \(A_\pm = 0\) in equation~\eqref{eq:m0_exact_sol}. They are~\cite{Nunez:2003eq,Myers:2007we}
\begin{equation}
    \w = 2 \pi T (n+1) (\pm 1 - i), \qquad n = 0, 1, 2, \cdots \; .
\end{equation}
Notice that these modes all collapse to \(\w=0\) in the limit \(T \to 0\). This is because the fluctuation equations are independent of \(b\) when \(m=0\), and thus when \(m=T=0\) there is no dimensionful scale that can set the spacing between the quasinormal modes.

When \(m \neq 0\), there can be quasinormal modes even zero temperature. We determine them numerically using a pseudospectral method, see for example refs.~\cite{boyd,Jansen:2017oag}. In this method, one approximates the solutions to the fluctuation equations (after removing the singular behaviour in the IR via a field redefinition) as a sum of the first \(K\) Chebyshev polynomials of the first kind. The fluctuation equations are then evaluated on an \(N\)-point grid, thus approximating the differential equation as an \(K \times K\) matrix equation, from which the quasinormal modes may be extracted as the eigenvalues. Details of our implementation of this method may be found in appendix~\ref{app:qnm}. To check convergence of our numerics, we evaluate the quasinormal modes for two different values of \(K\), keeping only the modes that agree following the algorithm in ref.~\cite{boyd}.

\begin{figure}[t]
    \begin{subfigure}{\textwidth}
    \raisebox{-0.5\height}{\includegraphics{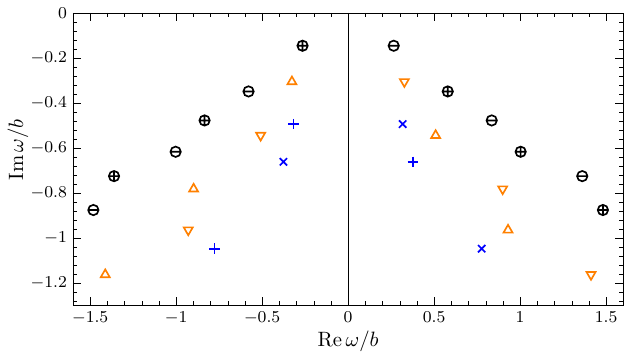}}
    \hfill
    \raisebox{-0.5\height}{\includegraphics{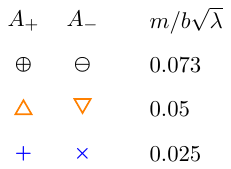}}
    \caption{Poles in \(\s_{xx}\) and \(\s_{xy}\).}
    \end{subfigure}
    \begin{subfigure}{\textwidth}
        \raisebox{-0.5\height} {\includegraphics{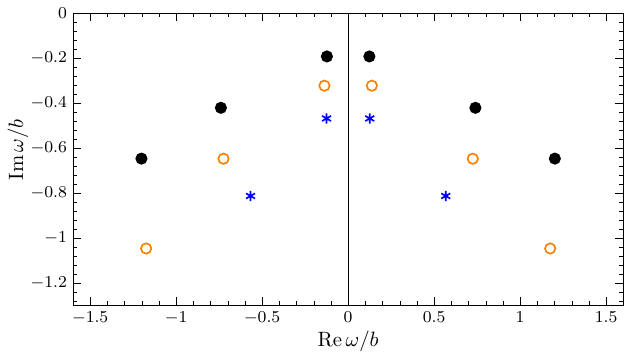}}
        \hfill
        \raisebox{-0.5\height}{\includegraphics{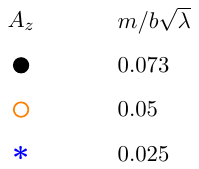}}
        \caption{Poles in \(\s_{zz}\).}
        \end{subfigure}
    \caption{Poles of the conductivities at zero temperature and sample values of \(m/b\sqrt{\l}\), in the complex frequency plane in units of \(b\). The top row shows poles in \(\s_{xx}\) and \(\s_{xy}\), which are dual to quasinormal modes of \(A_+\) and \(A_-\). The poles corresponding to quasinormal modes of \(A_+\) or \(A_-\) are indicated by different symbols, as shown in the legend. The bottom row shows poles in \(\s_{zz}\), dual to quasinormal modes of \(A_z\). These poles were computed using the pseudospectral method described in appendix~\ref{app:qnm}, using grid sizes \(K=90\) and \(K=100\). In addition to the poles plotted in this figure, we also see poles on the negative imaginary axis, that appear to coalesce to form a branch cut as \(N\) is increased, see figure~\ref{fig:branch_cut}.}
    \label{fig:T0_qnm_complex_plane}
\end{figure}

We will only compute the quasinormal modes for \(T=0\), leaving their determination at non-zero temperature to future work. The results of our numerical computation of the quasinormal modes for \(T=0\) and different sample values of \(m/b\sqrt{\l}\) are plotted in figure~\ref{fig:T0_qnm_complex_plane}. The modes were computed using a pseudospectral method, with grid sizes of \(K=90\) and \(K=100\). As anticipated, in both channels we see quasinormal modes that are relatively close to the real frequency axis for the largest mass plotted, \(m = 0.073 \, b \sqrt{\l}\), close to the phase transition. Decreasing \(m/b\sqrt{\l}\), the modes move away from the real frequency axis. 

As visible in figure~\ref{fig:T0_qnm_complex_plane}, the spectrum of quasinormal modes of \(A_z\) is symmetric under reflection through the imaginary frequency axis, i.e. under \(\w \to - \w^*\). This occurs because the equation of motion~\eqref{eq:A_z_eom} maps to the complex conjugate of itself under \(\w \to - \w^*\). Thus if \(A_z(\w;r)\) is a normalisable solution that obeys ingoing boundary conditions, then so is \(A_z(-\w^*;r)\).

On the other hand, the spectra of quasinormal modes of \(A_+\) and \(A_-\) are not invariant under reflection through the imaginary axis, but rather are mapped into each other. This occurs because the equations of motion~\eqref{eq:A_plus_minus_eom} for \(A_\pm\) are not mapped into (the complex conjugate of) themselves under the substitution \(\w \to - \w^*\) due to the presence of the term proportional to \(\w b\). This occurs because non-zero \(b\) breaks time reversal symmetry. Instead, the equation of motion for \(A_\pm\) maps into the equation of motion for \(A_\mp\) under \(\w \to - \w^*\), hence why the spectra of \(A_+\) and \(A_-\) are mirror images of each other. The same phenomenon has appeared in previous calculations of quasinormal modes of a holographic WSM~\cite{Rai:2024bnr}.

\begin{figure}[!t]
    \begin{center}
        \includegraphics{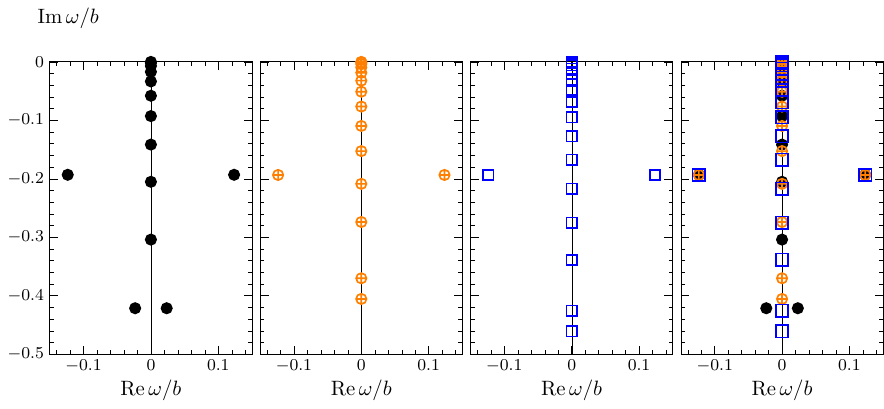}
    \end{center}
    \caption{Quasinormal modes of \(A_z\) close to the imaginary axis, at zero temperature and for \(m =0.073 \, b \sqrt{\l}\). The first three panels show, from left to right, the results of the computation for grid sizes \(K=60\) (black dots), \(K=80\) (orange \(\oplus\)), and \(K=100\) (blue squares). The right-hand panel shows these results superimposed upon each other, exhibiting the coalescence of modes on the imaginary axis as \(K\) is increased. There are also two modes at \(\w/b \approx \pm 0.12-0.19 \, i\) that do not move as \(K\) is changed. These modes are poles of \(\s_{zz}\), and are two of the modes plotted in figure~\ref{fig:T0_qnm_complex_plane}.}
    \label{fig:branch_cut}
\end{figure}

In addition to the modes plotted in figure~\ref{fig:T0_qnm_complex_plane}, we see many modes on the negative imaginary frequency axis. The density of these modes increases with increasing grid size, as plotted in figure~\ref{fig:branch_cut}, where we show the quasinormal modes of \(A_z\) for \(m = 0.073 \, b \sqrt{\l}\), calculated for three different values of the grid size; \(K=60\) (black dots), \(K=80\) (orange \(\oplus\)), and \(K=100\) (blue squares). The modes on or close to the imaginary axis move with \(N\), becoming denser as \(K\) is increased. Also visible in figure~\ref{fig:branch_cut} are two of the modes plotted in figure~\ref{fig:T0_qnm_complex_plane}, which do not move with \(K\). We find that the quasinormal modes of \(A_\pm\) behave qualitatively similarly to those of \(A_z\).

Modes coalescing on the imaginary axis in this way may be interpreted as evidence of the presence of a branch cut between \(\omega=0\) and infinity~\cite{Denef:2009yy,Edalati:2010hk,Edalati:2010pn,PremKumar:2020cxl}. Approximating the fluctuation equations as \(K \times K\) matrix equations splits a branch cut into discrete poles, which should coalesce in the limit \(K \to \infty\). Conversely, the isolated modes plotted in figure~\ref{fig:branch_cut} that do not move with increasing \(K\) should correspond to poles of the conductivity.

As discussed above, when \(m=T=0\) there are no quasinormal modes; the only possible singularities in the conductivities are branch cuts. The results in figure~\ref{fig:T0_qnm_complex_plane} suggest that the way the quasinormal modes disappear as \(m \to 0\) is by moving deeper into the complex plane, \(\Im \w \stackrel{?}{\to} \infty\). However, it is difficult to confirm this as we find that our pseudospectral determination of the quasinormal modes does not converge well for small~masses.\footnote{This is likely related to the fact that the magnitudes of the imaginary parts of the quasinormal modes are becoming larger. Quasinormal modes with large negative imaginary parts are highly unstable to perturbations, see e.g. ref.~\cite{Arean:2024afl}, so a small error in the numerical evaluation of the fluctuation equations will lead to a large change in the spectrum.}

The presence of quasinormal modes close to the real frequency axis may be roughly connected to the structure of the phase transition. As discussed in section~\ref{sec:review}, for the WSM phase/black hole embeddings at zero temperature, different values of \(m/b\sqrt{\l}\) correspond to different values of the IR parameter \(\h\) appearing in equation~\eqref{eq:bh_embedding_ir_T0}. As \(\h\) is increased, the black hole embeddings come closer and closer to the critical embedding. Similarly, the Minkowski embeddings approach the critical embedding as the parameter \(R_0\) in equation~\eqref{eq:minkowski_ir} is decreased to zero.

As the critical embedding is approached from either side, we expect the spectra of the black hole and Minkowski embeddings to become the same. Since Minkowski embeddings have normal modes, with vanishing imaginary part, this implies that the imaginary parts of the quasinormal modes of the black hole embeddings should vanish as \(\h \to \infty\).\footnote{See refs.~\cite{Hoyos-Badajoz:2006dzi,Kaminski:2009ce} for results for the near-critical limit for \(b=0\) and non-zero temperature.} Thus, the existence of modes with small imaginary parts for \(m/b \sqrt{\l}\) close to the phase transition can perhaps be understood as being due to black hole embeddings close to the phase transition being ``close to'' the critical embedding in some sense.

\begin{figure}[!htbp]
    \begin{center}
        \includegraphics{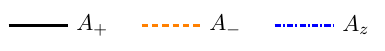}
    \end{center}
    \vspace{-1.5em}
    \begin{subfigure}{0.5\textwidth}
        \includegraphics{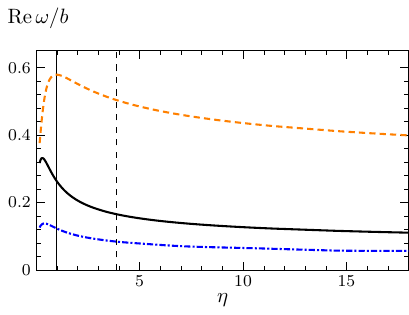}
    \end{subfigure}\begin{subfigure}{0.5\textwidth}
        \includegraphics{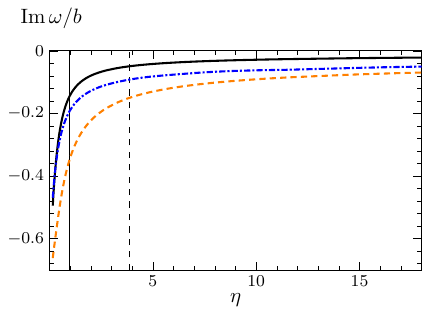}
    \end{subfigure}
    \begin{subfigure}{0.5\textwidth}
        \includegraphics{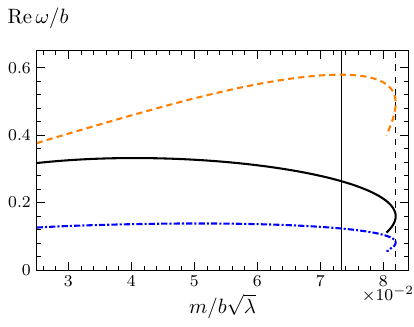}
    \end{subfigure}\begin{subfigure}{0.5\textwidth}
        \includegraphics{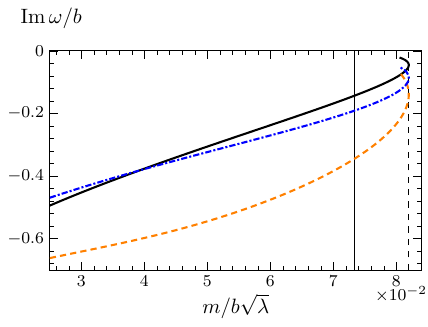}
    \end{subfigure}
    \caption{The real and imaginary parts of the quasinormal modes with positive real part closest to origin in the complex plane for each gauge field fluctuation, plotted as functions of the IR parameter \(\h\) (top row) and \(m/b\sqrt{\l}\) (bottom row). The solid vertical lines in each plot show the location of the first-order phase transition, at \(m \approx 0.0733 \, b \sqrt{\l}\). The dashed vertical lines show the location of the maximum mass of the black hole embeddings, at \(m \approx 0.0820 \, b \sqrt{\l}\). Black hole embeddings approach the critical embedding for large \(\h\). Consequently, the imaginary parts of the frequencies of the quasinormal modes approach zero as \(\h\) is increased. The modes in this figure were computed using grid sizes \(K=90\) and \(K=100\).}
    \label{fig:qnm_re_im}
\end{figure}

As evidence for this claim, in figure~\ref{fig:qnm_re_im} we plot the real and imaginary parts of the modes closest to the real frequency axis in each of the gauge field fluctuation channels. Due to the symmetry of the spectra under reflections in the imaginary axis, we only show modes with positive real part. The solid black, dashed orange, and dot-dashed blue curves in the figure show quasinormal modes of \(A_+\), \(A_-\), and \(A_z\), respectively.

The top row in the figure shows the frequencies as functions of \(\h\). We see that the imaginary parts of the modes indeed appear to decrease to zero as \(\h\) is increased, matching the expectation explained above. Unfortunately, the numerical computations become more difficult at large \(\h\), so from our results it is difficult to tell whether the real parts of the frequencies vanish very slowly or asymptote to some non-zero value at large \(\h\).

The second row of figure~\ref{fig:qnm_re_im} shows the frequencies of the same modes but plotted as functions of \(m/b\sqrt{\l}\), rather than \(\h\). Note that the frequencies ``turn around'' at the maximal value of \(m/b\sqrt{\l} \approx 0.0819 \) because the mass is a non-injective function of \(\h\). As a guide to the translation between \(\h\) and \(m/b\sqrt{\l}\), in the figures we have added solid vertical lines, showing the location of the first order phase transition at \(m/b\sqrt{\l} \approx 0.0733\), corresponding to \(\h \approx 0.989\), and dashed vertical lines showing the maximal value of \(m/b\sqrt{\l}\), corresponding to \(\h \approx 3.89\). Notice that although the phase transition occurs for \(\h\) close to one and approaching the critical embedding corresponds to the limit \(\h \to \infty\), the values of \(m/b\sqrt{\l}\) at \(\h \approx 1\) and \(\h \gg 1\) differ only by around 10\%, so that embeddings in these two regimes are somewhat similar except in the deep IR \(r \to 0\).

The imaginary parts of quasinormal modes of zero-momentum gauge field fluctuations have also been found to become small close to the phase transition in a bottom-up holographic WSM model~\cite{Rai:2024bnr}. However, the phase transition in this model is second-order, so the mechanism for the smallness of \(\Im \w\) appears to be different.

\section{Discussion}
\label{sec:discussion}

In this work we have computed the AC conductivities in the holographic model of a WSM based on flavour D7-branes in a D3-brane background. We found that at zero or low temperature and for masses close to the transition to an insulating phase, the longitudinal conductivity \(\s_{xx}\) for directions orthogonal to the applied axial field exhibits a peak at frequencies \(\w \sim 0.1 \, b\). Computing the poles of this conductivity, we argued that this peak arises from a pole that comes close to the real frequency axis due to the phase transition. We also found that the Hall conductivity and the longitudinal conductivity exhibit troughs as functions of frequency in roughly similar regimes.

There are a number of possible directions for future research. For example, our results for the real parts of the longitudinal conductivities scale linearly with frequency at both small and large frequency. The authors of ref.~\cite{Grignani:2016wyz} have argued that experimental results for WSM conductivities are better approximated by conductivities scaling at small frequencies as \(\Re \s_{xx} \propto \w^\b\) and \(\Re \s_{zz} \propto \w^{2-\b}\) with exponent \(\b \approx 0.14\). It would be interesting to try to construct a top-down holographic model that reproduces this exponent.

There are also many other channels in which one could study the holographic Green's functions and quasinormal modes of the flavour brane WSM model, for example, for example one can study fluctuations of the embedding scalar \(R\) to compute the two point function of the QFT's scalar operator \(\cO_m\), and its poles. One motivation to perform a more thorough analysis of the fluctuations of our model is to check the linear stability of D7-brane embeddings dual to the WSM phase.

The fluctuations of \(R\) and of the gauge field could also be studied at non-zero momentum. Due to the anisotropy introduced by the applied axial gauge field \(\vec{A}_5\), the Green's functions will depend differently on momentum parallel and perpendicular to \(\vec{A}_5\). This leads to a richer hydrodynamic description than for Lorentz-invariant systems (see e.g. ref.~\cite{Landsteiner:2019kxb}), which could be compared to the holographic Green's function calculations at small frequencies and momentum. This would allow the determination of various transport coefficients, beyond the conductivities considered in this work.

As mentioned in the introduction, WSMs with a boundary have current-carrying surface states. Ref.~\cite{Ammon:2016mwa} demonstrated the existence of surface currents in a bottom-up holographic WSM model, by constructing an interface between the WSM and a topologically trivial semimetal. One could repeat their analysis in the flavour brane model. This would involve the construction of an interface between the WSM and insulating phases of the flavour brane model, and determining whether one finds a current localised to said interface. It may also be interesting to study fermionic spectral functions in the presence of such an interface.

\section*{Acknowledgments}

We would like to thank Mahdi Atashi, Kazem Bitaghsir Fadafan, Karl Landsteiner, and  Andy O'Bannon for useful discussions. We would particularly like to thank Kazem Bitaghsir Fadafan and Andy O'Bannon for comments on a draft of this manuscript. The work of R.R. was supported by the European Union's Horizon Europe research and innovation program under Marie Sklodowska-Curie Grant Agreement No.~101104286. Nordita is supported in part by Nordforsk.

\appendix

\section{Formulas for AC conductivities}
\label{app:conductivities}

Starting from the ansatz~\eqref{eq:D7_ansatz}, we add small fluctuations of the gauge field components \(A_i\), depending on \(t\) and \(r\). Expanding the D7-brane action~\eqref{eq:D7_action} in powers of the gauge field fluctuations, the part that is quadratic in the \(A_i(t,r)\) is
\begin{equation}\begin{aligned}
    S^{(2)}_\mathrm{D7} = \frac{\cN}{2} \int \diff t \, \diff r \, \frac{r^3}{\r^6} \sqrt{h} \sqrt{\frac{\r^4 h + b^2 R^2}{1 + R'^2}} \biggl[&
        \frac{1 + R'^2}{f} \le(\dot{A}_x^2 + \dot{A}_y^2 + \frac{\r^4 h}{\r^4 h + b^2 R^2} \dot{A}_z^2 \ri)
        \\
        & - \r^4 \frac{f}{h} \le(A_x'^2 + A_y'^2 + \frac{\r^4 h}{\r^4 h + b^2 R^2} A_z'^2 \ri)
        \\
        & + 2 b \frac{r^4}{\r^4} (\dot{A}_y A_x' - \dot{A}_x A_y')
    \biggr],
\end{aligned}\end{equation}
where  \(\cN = 8 \pi^4 N_f L^4 T_\mathrm{D7} \int \diff^3 \vec{x} = \frac{N_f N_c}{4\pi^2} \int \diff^3 \vec{x}\), dots denote derivatives with respect to \(t\), and primes denote derivatives with respect to \(r\). On shell, we can use the equations of motion for the \(A_i\) to reduce this action to a boundary term which we denote \(S^{(2)\star}_\mathrm{D7}\), evaluated on a cutoff surface at large \(r = r_c\),
\begin{equation} \label{eq:gauge_field_on_shell_action_intermediate}
    S^{(2)\star}_\mathrm{D7} = - \frac{\cN}{2} \int \diff t \le[\frac{r^3f}{\r^2\sqrt{h}} \le( A_x A_x' + A_y A_y' + \frac{\r^4 h}{\r^4 h + b^2 R^2} A_z A_z' \ri) - b \frac{r^4}{\r^4} (A_x \dot{A}_y - A_y \dot{A}_x)\ri]_{r_c},
\end{equation}
We then Fourier transform with respect to time, defining \(A_i(t,r) = \int \frac{\diff \w}{2 \pi} e^{-i \w t} A_i (\w;r)\). The Fourier coefficients have the near-boundary expansions
\begin{equation}
    A_i(\w;r) = A_i^{(0)}(\w) \le[1 + \frac{\w^2}{4r^2} \log(r^2/\w^2)\ri] + \frac{A_i^{(2)}(\w)}{r^2} + O(r^{-4} \log r),
\end{equation}
where \(A_i^{(0)}(\w)\) and \(A_i^{(2)}(\w)\) are integration coefficients. Substituting these expansions into the on-shell action~\eqref{eq:gauge_field_on_shell_action_intermediate} we find
\begin{equation}\begin{aligned}
    S^{(2)\star}_\mathrm{D7} = \cN\int \diff t \biggl[
        A_i^{(0)}(-\w) A_i^{(2)}&(\w) - i \w b A_x^{(0)}(-\w) A_y^{(0)}(\w)
        \\
        &- \frac{\w^2}{4} \le(1 - \log(r_c^2/\w^2)\ri) A_i^{(0)}(-\w) A_i^{(0)}(\w) 
    \biggr].
\end{aligned}\end{equation}
There is a term that diverges logarithmically as we remove the cutoff by sending \(r_c \to \infty\). To obtain a well-defined variational principle and finite results for observables we must remove this divergent term with a counterterm \(S_\mathrm{ct}\), a boundary term evaluated at \(r=r_c\)~\cite{deHaro:2000vlm,Karch:2005ms}. The necessary counterterm is~\cite{Hoyos:2011us}
\begin{equation}\begin{aligned}
    S_\mathrm{ct} &= \frac{\cN}{8} \int \diff^4 x \sqrt{-\g} \, F^{\m\n} \log(r_c^2/\Box)F_{\m\n}
    \\
    &=- \frac{\cN}{4} \int \frac{\diff \w}{2\pi} \w^2 A_i^{(0)}(-\w) A_i^{(0)}(\w) \log(r_c^2/\w^2).
\end{aligned}\end{equation}
Adding this counterterm, the renormalised on-shell action \(S_\mathrm{ren}^\star = S^{(2)\star}_\mathrm{D7} + S_\mathrm{ct}\) for the gauge field fluctuations becomes
\begin{equation}
    S_\mathrm{ren}^\star = \cN \int \frac{\diff \w}{2\pi} \le[
        A_i^{(0)}(-\w) A_i^{(2)}(\w) - \frac{\w^2}{4} A_i^{(0)}(-\w) A_i^{(0)}(\w) - i \w b A_x^{(0)}(-\w) A_y^{(0)}(\w)
    \ri].
\end{equation}
Applying the Lorentzian correlator prescription of refs.~\cite{Son:2002sd,Herzog:2002pc,Kaminski:2009dh} we obtain expressions for the retarded Green's functions of the vector current in the dual QFT, at frequency \(\w\) and vanishing momentum,
\begin{equation}\begin{aligned} \label{eq:greens_function_formula}
    G_{J^x J^x}(\w, k = 0 ) &= \frac{N_f N_c}{4 \pi^2} \le(2 \frac{A_x^{(2)}(\w)}{A_x^{(0)}(\w)} - \frac{\w^2}{2} \ri),
    \\
    G_{J^x J^y}(\w, k = 0 ) &= \frac{N_f N_c}{4 \pi^2} \le(2 \frac{A_y^{(2)}(\w)}{A_x^{(0)}(\w)} + i \w b \ri),
    \\
    G_{J^z J^z}(\w,k=0) &= \frac{N_f N_c}{4 \pi^2} \le(2 \frac{A_z^{(2)}(\w)}{A_z^{(0)}(\w)} - \frac{\w^2}{2} \ri),
\end{aligned}\end{equation}
where all of these expressions are to be evaluated with \(A_n(\w;r)\) satisfying ingoing boundary conditions at the horizon (or at \(r=0\) at zero temperature). Further, the expressions for \(G_{J^x J^x}(\w,k=0)\) and \(G_{J^x J^y}(\w,k=0)\) are to be evaluated with boundary conditions such that \(A_y^{(0)}=0\). Substituting equation~\eqref{eq:greens_function_formula} into the Kubo formula~\eqref{eq:conductivity_kubo_formula} and using \(A_\pm = A_x \pm i A_y\) to replace \(A_{x,y}^{(2)}\) and \(A_x^{(0)}\) with \(A_\pm^{(2)}\) and \(A_\pm^{(0)}\), one obtains the formulas in equation~\eqref{eq:sigma_formula}.

\section{Numerical details}

In this appendix we summarise the numerical methods used to generate the results shown in the main text. For both the conductivities and the quasinormal modes we must first numerically solve the embedding equation~\eqref{eq:embedding_equation} for \(R(r)\), subject to the boundary conditions that \(R(r \to \infty) = M\) and that \(R(r)\) behaves as a black hole embedding in the deep IR. The solution for \(R(r)\) is then substituted into the fluctuation equations~\eqref{eq:fluctuation_equations}, which must in turn be solved numerically.

\subsection{Conductivities}
\label{app:pseudospectral}

To obtain the conductivities we solve both the embedding equation~\eqref{eq:embedding_equation} and the fluctuation equations~\eqref{eq:fluctuation_equations} using a pseudospectral collocation method based on Chebyshev polynomials. See ref.~\cite{boyd} for more technical details. The implementation of this method is slightly different for zero versus non-zero temperature, so we describe each case separately.

\subsubsection{Zero temperature}
\label{app:pseudospectral_zero_T}

To implement the pseudospectral method we define a new, dimensionless radial coordinate \(\fr = (r-b)/(r+b)\) which takes values in the range \(\fr \in [-1,1]\), with the Poincar\'e horizon at \(\fr = -1\) and boundary at \(\fr = 1\). We solve the embedding equation by approximating \(R\) as a sum of Chebyshev polynomials of the first kind \(T_k\),
\begin{equation} \label{eq:chebyshev_sum}
    R(\fr) = \sum_{k=0}^{N-1} c_k T_k(\fr)
\end{equation}
for some large positive integer \(K\). We fix the coefficients \(c_k\) by solving equation~\eqref{eq:embedding_equation} (transformed to the \(\fr\)) coordinate on the interior points of the Gauss--Lobatto grid,
\begin{equation}
    \fr_k = \cos\le(\frac{\pi k}{K-1}\ri), \qquad k = 1, 2, \cdots, K-2,
\end{equation}
supplemented with the boundary conditions \(R(\fr = -1) =0\) and \(R(\fr=1) = M\) for the desired value of \(M\).

It turns out to be convenient to solve not for the coefficients \(c_k\)~\eqref{eq:chebyshev_sum}, but rather the values \(R(\fr_k)\) evaluated on the Gauss--Lobatto grid, which contain equiavalent information to the \(c_k\) (they are related by a discrete cosine transform). Differentiation identities for the Chebyshev polynomials may then be used to express \(R'(\fr_k)\) and \(R''(\fr_k)\) as linear combinations of \(R\) evaluated on the rest of the grid. This is conveniently implemented in Mathematica using the differentiation matrices from \texttt{NDSolve\`{}FiniteDifferenceDerivative} with the option \texttt{DifferenceOrder} \(\to\) \texttt{"Pseudospectral"}.

Since the equation of motion for \(R(\fr)\) is non-linear, we solve for the \(R(\fr_k)\) using the Newton procedure described in refs.~\cite{boyd,Krikun:2018ufr}. This requires an initial seed \(R_0(\fr_k)\) for the values of \(R\) on the grid points. We find that setting \(R_0(\fr_k) = 0\) works well.

Having obtained a numerical solution for \(R(\fr)\), we would then like to substitute this solution into the equations of motion for the gauge field fluctuations \(A_n(\w;r)\), which we would then solve by the same pseudospectral method to determine the near-boundary conditions \(A_n^{(0)}(\w)\) and \(A_n^{(2)}(\w)\), which in turn determine the conductivities through equation~\eqref{eq:sigma_formula}. However, we face two complications.

The first complication is that the ingoing boundary conditions~\eqref{eq:ingoing_boundary_conditions} obeyed by the fluctuations are non-analytic at the Poincar\'e horizon \(\fr =-1\), and therefore \(A_n(\w;r)\) cannot be well approximated by a sum of Chebyshev polynomials. To solve this we define new fields
\begin{equation}
    Z_n(\w;r) = \le(1 + \frac{b^2}{r^2} \ri)^{-1} e^{-i \w/r} A_n(\w;r),
\end{equation}
in terms of which the ingoing boundary conditions become \(Z_n(\w;0) = 0\). The \(Z_n\) are therefore much more amenable to pseudospectral treatment.

The second complication comes from the determination of \(A_n^{(2)}(\w)\). In the \(\fr\) coordinate, the near-boundary expansion of the rescaled fluctuation is
\begin{multline}
    Z_n(\w;\fr) = A_n^{(0)}(\w) \le[1 + \frac{i \w}{2b} (\fr -1) + \frac{\w^2}{8b^2}(1-\fr)^2 \log\le( \frac{2b}{|\w| (1-\fr)}\ri)\ri]
    \\
    + \frac{(1- \fr)^2}{8} \le[\frac{2 A_n^{(2)}(\w)}{b^2} - A_n^{(0)}(\w) \le(2 + \frac{2 i \w}{b} + \frac{\w^2}{b^2} \ri)\ri] + O\le((1-\fr)^3 \log(1-\fr)\ri) .
\end{multline}
There is a logarithmic term that is more important as \(\fr \to 1\) than the term containing \(A_n^{(2)}(\w)\), which makes it difficult to reliably determine \(A_n^{(2)}(\w)\) from an approximation of \(Z_n(\w;r)\) in terms of polynomials. One possible solution would be to explicitly subtract the logarithmic term from \(Z_n(\w;\fr)\), however we would still face the problem that \(A_n^{(2)}(\w)\) appears at a subleading order in the near-boundary expansion, so we would need to work at rather large precision to reliably extract \(A_n^{(2)}(\w)\), which would slow down numerics.

A more elegant solution is to define
\begin{equation}
    p_n(\w;\fr) = (1+\fr) \le[\p_{\fr}  - \frac{i \w}{2b} - \frac{\w^2}{4 b^2} (1-\fr) \log(1-\fr) - \frac{3 \w^2}{8b^2} (1-\fr)^2 \log(1-\fr) \ri] Z_n(\w;r).
\end{equation}
At leading order near the boundary we find
\begin{equation}
    p_n(\w;\fr) = (\fr-1)\frac{A_n^{(2)}(\w)}{b^2} - (1-\fr) A_n^{(0)} \le[1 + \frac{i \w}{b} + \frac{\w^2}{4 b^2} + \frac{\w^2}{2 b^2} \log\le(\frac{|\w|}{2b} \ri) \ri]+ O((1-\fr)^2).
\end{equation}
We recast the second-order equation from \(Z_n(\w;\fr)\) as a pair of coupled first-order equations for \((Z_n(\w;\fr),p_n(\w;\fr))\), which we then solve using the pseudospectral method, subject to the boundary condition \(Z_n(\w;1) = 1\), corresponding to the choice \(A_n^{(0)}(\w)=1\),\footnote{We are free to choose any non-zero value of \(A_n^{(0)}(\w;r)\) since the fluctuation equations are linear, so given a solution with some non-zero \(A_n^{(0)}(\w;r)\), we can rescale that solution to obtain another solution with \(A_n^{(0)}(\w)=1\). The conductivities are proportional to \(A_n^{(2)}(\w)/A_n^{(0)}(\w)\) so are invariant under this rescaling.} and \(Z_n(\w;-1) = 0\). Differentiating the resulting solution for \(p_n(\w;\fr)\) at \(\fr=1\) we obtain \(A_n^{(2)}(\w)\). Following this procedure for each \(n \in (+,-,z)\), we then use the formulas in equation~\eqref{eq:sigma_formula} to obtain the conductivities \(\s_{xx}\), \(\s_{xy}\) and \(\s_{zz}\).

\subsubsection{Non-zero temperature}

At non-zero temperature we use the same pseudospectral method to solve for the embedding and fluctuations. Compared to the zero temperature case, the range of the radial coordinate \(r\) and the ingoing boundary conditions obeyed by the fluctuations are both different, so we will need to perform different variable redefinitions.

At a given temperature \(T\), corresponding to given horizon radius \(\r_h\), black hole embeddings are distinguished by \(r_h\), the value of \(r\) at which they meet the horizon. Thus the radial coordinate for these embeddings takes values in the range \(r \in [r_h,\infty)\). We find it convenient to replace \(r\) with \(\r = \sqrt{r^2 + R^2}\), which takes values in a range that is the same for all embeddings at a given temperature, \(\r \in [\r_h,\infty)\). We then map \(\r\) to the finite interval \(\fr \in [-1,1]\) by defining 
\begin{equation}
    \fr = 1 - \frac{2\r_h}{\r},
\end{equation}
where \(\fr = -1\) corresponds to the horizon, while \(\fr=1\) again corresponds to the conformal boundary. We also find it convenient to define a dimensionless variable \(u(\fr) =(1+\fr) R(\fr)/2b\), such that the IR boundary condition on becomes \(u(\fr=-1) = 0\).

In order to turn the ingoing boundary condition~\eqref{eq:ingoing_boundary_conditions} into regularity boundary conditions, we define new fluctuation variables
%
\begin{equation} \label{eq:finite_t_fluctuation_rescaling}
    Z_n(\w;\r) = \le(\frac{2 \r - i \sqrt{2} \sqrt{1 + \r^4/\r_h^4} }{2 \r + i \sqrt{2} \sqrt{1 + \r^4/\r_h^4}}\ri)^{\frac{(1-i) \w}{4\pi T}}
    \exp \le[\frac{i \w}{2 \pi T} \le(1 - \frac{\sqrt{2}\, \r / \r_h}{\sqrt{1 + \r^4/\r_h^4}} \ri) \ri] A_n(\w;\r).
\end{equation}
This is not the simplest possible rescaling that would make \(Z_n(\w;r)\) regular at the horizon. It contains additional factors that essentially amount to the exchanging the time coordinate \(t\) in equation~\eqref{eq:ads5_cross_s5} for the time coordinate of \ads[5]-Schwarzschild in regular coordinates.\footnote{See e.g. refs.~\cite{Arean:2024afl,Garcia-Farina:2024pdd} for a discussion of regular coordinates} These two time coordinates agree at the boundary, so the exchange does not affect the frequency dependence of the conductivities we compute. Adopting regular coordinates improves the convergence of pseudospectral computations of quasinormal modes.\footnote{We thank Karl Landsteiner for making us aware of this.} Similarly, we find that the additional rescalings in equation~\eqref{eq:finite_t_fluctuation_rescaling} improve the convergence of our pseudospectral determination of the conductivities.

Finally, in order to reliably extract \(A_n^{(2)}(\w)\) we define
\begin{equation}
    \Pi_n(\w;\fr) = \frac{1+\fr}{2} \le[\p_\fr - \frac{\w^2}{2 \pi^2 T^2} (1-\fr) \log(1-\fr) \ri] Z_n(\w;\fr),
\end{equation}
which behaves near the boundary as
\begin{multline}
    \Pi_n(\w;r) \approx \exp\le[-\frac{i\w}{\pi T} \le(1 + \frac{(1-i)\pi}{4}\ri)\ri] 
    \\
    \times \le[\frac{A_n^{(2)}(\w)}{\pi^2 T^2} -\frac{\w^2 A_n^{(0)}(\w)}{4 \pi^2 T^2} \le(1  +\log \le(\w^2 /2 \pi^2 T^2 \ri) \ri) \ri](\fr-1) + O\le( (\fr-1)^2\ri).
\end{multline}
We then apply the same approach as for zero temperature, recasting the equations of motion for the fluctuation \(A_n(\w;\fr)\) as a pair of coupled first-order equations for \(Z_n(\w;\fr)\) and \(\Pi_n(\w;\fr)\), subject to the boundary conditions \(Z_n(\w;1) = 1\) and \(Z_n(\w;-1) = 0\), which we solve pseudospectrally.

\subsection{Quasinormal modes}
\label{app:qnm}

To compute the zero-temperature quasinormal modes using the pseudospectral method, we employ the same radial coordinate \(\fr\) as in section~\ref{app:pseudospectral_zero_T}. The quasinormal mode boundary conditions are that \(A_n(\w;r)\) is ingoing at \(r=0\) (corresponding to \(\fr=-1\)) and \(A_n^{(0)}=0\). In terms of the dependent variable \(Z_n\) defined in section~\ref{app:pseudospectral_zero_T}, these boundary conditions are \(Z_n(\w;\fr=\pm1) = 0\).

Near \(\fr=-1\), solutions obeying ingoing boundary conditions behave as \(Z_n(\w;\fr) \propto (1 +\fr)^{3/2}\). It turns out to improve the convergence of the pseudospectral quasinormal mode method if we redefine the dependent variable to remove this non-analytic behaviour. Specifically, we define
\begin{equation}
    \cZ_n(\w;\fr) = \frac{Z_n(\w;\fr)}{(1+\fr)^{3/2}},
\end{equation}
such that ingoing boundary conditions correspond to \(\cZ_n(\w;r)\) remaining finite in the limit \(\fr \to -1\).

To determine the quasinormal modes, similar to section~\ref{app:pseudospectral_zero_T} we approximate \(\cZ_n(\w;r)\) as a sum of the first \(K\) Chebyshev polynomials of the first kind, which we then use to evaluate the equation of motion for \(\cZ_n(\w;\fr)\) on the interior points of the \(K\)-point Gauss--Lobatto grid. We supplement this with the normalisibility condition \(\cZ(\w;1) =0\), and the \(\fr \to -1\) limit of the equation of motion, which imposes that
\begin{equation}
    16 i \w \p_\fr \cZ_n(\w;\fr)|_{\fr=-1} = 3(4 i \w -  b) \cZ_n(\w;-1).
\end{equation}
This procedure approximates the equation of motion for \(\cZ_n\) as a matrix equation of the forms
\begin{equation} \label{eq:qnm_equation}
    \le(P + \w^2 Q + \w R\ri) \vec{\cZ}_n = 0,
\end{equation}
where the components of \(\vec{\cZ}_n\) are \(\cZ_n(\w;\fr_i)\), i.e. the dependent variable evaluated on the Gauss--Lobatto grid, while \(P\), \(Q\), and \(R\) are \(K \times K \) matrices. The quasinormal modes are the generalised eigenvalues \(\w\) of equation~\eqref{eq:qnm_equation}, which we determine using \texttt{Eigenvalues} in \texttt{Mathematica}.

\section{Additional plots}
\label{app:plots}

\begin{figure}[!htbp]
    \begin{center}\includegraphics{figs/conductivity_legend.pdf}\end{center}\vspace{-12pt}
    \begin{subfigure}{\textwidth}
        \includegraphics{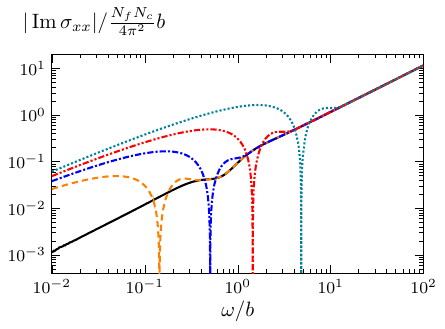}
        \includegraphics{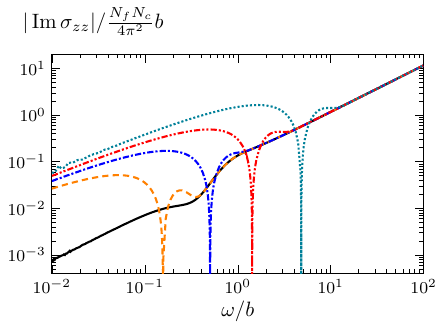}
        \caption{\(m = 0.025 \,  b \sqrt{\l}\)}
    \end{subfigure}
    \begin{subfigure}{\textwidth}
        \includegraphics{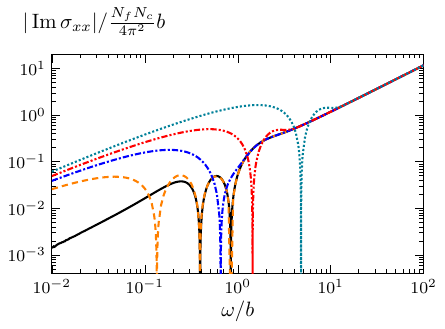}
        \includegraphics{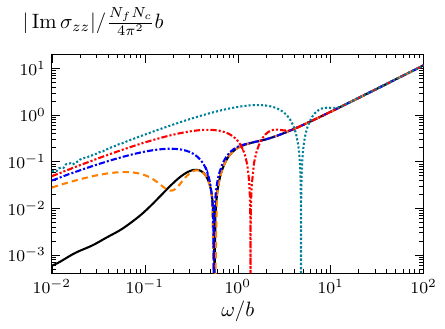}
        \caption{\(m = 0.05 \, b \sqrt{\l}\)}
    \end{subfigure}
    \begin{subfigure}{\textwidth}
        \includegraphics{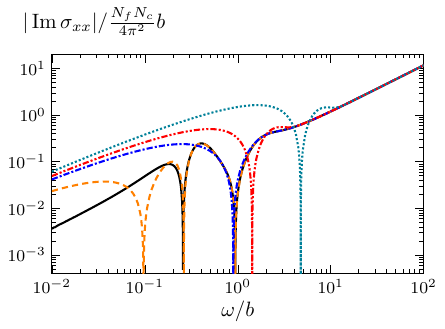}
        \includegraphics{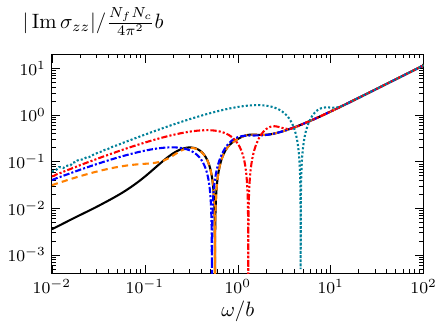}
        \caption{\(m = 0.073 \, b \sqrt{\l}\)}
    \end{subfigure}
    \caption{Imaginary parts of the longitudinal AC conductivities as functions of frequency at non-zero temperature, at various values of \(T/b\) and \(m/b\sqrt{\l}\). The real parts of these conductivities are plotted in figure~\ref{fig:conductivity_real}.  We plot the absolute values of the imaginary values to be able to show them on logarithmic axes. The signs can be inferred from the information that \(\Im \s_{xx}\), \(\Im \s_{xy}\) and \(\Im \s_{zz}\) are all negative at large frequencies.}
    \label{fig:conductivity_imaginary}
\end{figure}

\begin{figure}[!htbp]
    \begin{center}\includegraphics{figs/conductivity_legend.pdf}\end{center}\vspace{-12pt}
    \begin{subfigure}{0.5\textwidth}
        \includegraphics{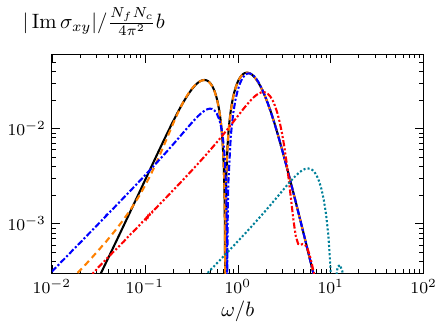}
        \caption{\(m = 0.025 \,  b \sqrt{\l}\)}
    \end{subfigure}\begin{subfigure}{0.5\textwidth}
        \includegraphics{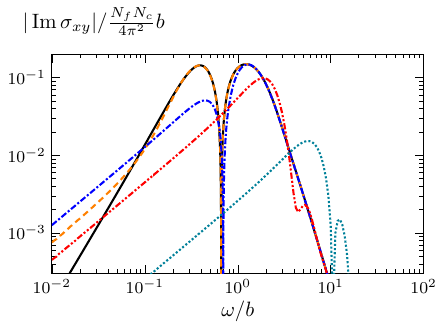}
        \caption{\(m = 0.05 \, b \sqrt{\l}\)}
    \end{subfigure}
    \begin{subfigure}{\textwidth}
        \begin{center}\includegraphics{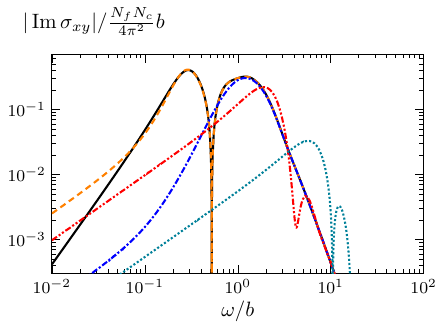}\end{center}
        \caption{\(m = 0.073 \, b \sqrt{\l}\)}
    \end{subfigure}
    \caption{Imaginary part of the AC Hall conductivity \(\s_{xy}\) as a function of frequency at non-zero temperature, at various values of \(T/b\) and \(m/b\sqrt{\l}\). The real part of this conductivity is plotted in figure~\ref{fig:hall_conductivity_real}.  We plot the absolute values of the imaginary values to be able to show them on logarithmic axes. The signs can be inferred from the information that \(\Im \s_{xx}\), \(\Im \s_{xy}\) and \(\Im \s_{zz}\) are all negative at large frequencies.}
    \label{fig:hall_conductivity_imaginary}
\end{figure}

This appendix contains plots of the imaginary parts of the AC conductivities as functions of frequency, for non-zero temperature and various values of \(m/b\sqrt{\l}\). Figure~\ref{fig:conductivity_imaginary} shows the imaginary part of the longitudinal conductivities \(\s_{xx}\) and \(\s_{zz}\), whose real parts we plotted in figure~\ref{fig:conductivity_real}. Figure~\ref{fig:hall_conductivity_imaginary} shows the imaginary part of the Hall conductivity \(\s_{xy}\), whose real part we plotted in figure~\ref{fig:hall_conductivity_real}.

\clearpage

\bibliographystyle{JHEP}
\bibliography{wsm}

\end{document}